\documentclass[twocolumn]{aastex63}

\usepackage[utf8]{inputenc}
\usepackage{enumitem}
\usepackage{amsmath}
\usepackage{graphicx}
\usepackage{hyperref}
\usepackage{color}
\hypersetup{colorlinks=true, citecolor=cyan, urlcolor = blue, linkcolor = blue}
\newcommand{\rev}[1]{\textcolor{black}{{#1}}}
\newcommand{\revsjv}[1]{\textcolor{black}{{#1}}}
\newcommand{\revtwo}[1]{\textcolor{black}{{#1}}}

\newcommand{\fgw}{f_{\rm GW}}

\newcommand{\lowesth}[1]{$(6.82 \pm 0.35) \times 10^{-15}$}
\newcommand{\lowesthonly}[1]{$(9.11 \pm 0.10) \times 10^{-15}$}
\newcommand{\bestfreq}[1]{$7.65\times10^{-9}$~Hz}
\newcommand{\lowesthold}[1]{$(7.33 \pm 0.29) \times 10^{-15}$}

\newcommand{\bestpix}[1]{$(2.66 \pm 0.15) \times 10^{-15}$}
\newcommand{\worstpix}[1]{$(1.12 \pm 0.05) \times 10^{-14}$}

\newcommand{\farthestD}[1]{35.0 Mpc}
\newcommand{\farthestF}[1]{$3.82\times10^{-8}$ Hz}
\newcommand{\senseD}[1]{33.9 Mpc}

\newcommand{\farthestDmap}[1]{86.7 Mpc}
\newcommand{\shortestDmap}[1]{20.5 Mpc}

\newcommand{\farthestDmapTen}[1]{4.02 Gpc}
\newcommand{\shortestDmapTen}[1]{0.95 Gpc}

\newcommand{\bftargetcrn}[1]{$0.67 \pm 0.01$}
\newcommand{\bftargetcw}[1]{$0.70 \pm 0.02$}

\newcommand{\ultargetcrn}[1]{$\mathcal{M} <(1.41 \pm 0.02) \times 10^9 M_\odot$}
\newcommand{\ultargetcw}[1]{$\mathcal{M} <(1.34 \pm 0.01) \times 10^9 M_\odot$}

\newcommand{\ultargetold}[1]{$1.65 \times 10^9 M_\odot$}

\newcommand{\improvement}[1]{$2.4 \times 10^8 M_\odot$}
\newcommand{\improvementFactor}[1]{$1.2$}

\newcommand{\ultargetcrnH}[1]{$1.90 \times 10^{-14}$}
\newcommand{\ultargetcwH}[1]{$1.74 \times 10^{-14}$}

\newcommand{\ulAllSkyFreqCRN}[1]{$3.56 \times 10^{-14}$}
\newcommand{\ulAllSkyFreqOnly}[1]{$3.82 \times 10^{-14}$}

\newcommand{\facCRN}[1]{$1.88$}
\newcommand{\facOnly}[1]{$2.20$}

\shorttitle{NANOGrav 12.5-year Continuous Wave Limits}
\shortauthors{The NANOGrav Collaboration}

\received{2023 January 9}
\accepted{2023 June 6}

\begin{document}
\title{The NANOGrav 12.5-year Data Set: Bayesian Limits on Gravitational Waves from Individual Supermassive Black Hole Binaries}

\correspondingauthor{Caitlin A. Witt$^{\color{magenta}\S}$}
\email{caitlin.witt@nanograv.org}

% \author{NANOGrav}
% \affiliation{everywhere}
% DO NOT EDIT THIS FILE. EDITS WILL BE OVERWRITTEN.
% AUTO-GENERATED WITH make-aastex62-author-list.py
% FROM author_list_12yr_cw.txt, author_affil_and_orcid.txt, AND affil.txt
\author{Zaven Arzoumanian}
\affiliation{X-Ray Astrophysics Laboratory, NASA Goddard Space Flight Center, Code 662, Greenbelt, MD 20771, USA}
\author[0000-0003-2745-753X]{Paul T. Baker}
\affiliation{Department of Physics and Astronomy, Widener University, One University Place, Chester, PA 19013, USA}
\author{Laura Blecha}
\affiliation{Department of Physics, University of Florida, 2001 Museum Rd., Gainesville, FL 32611, USA}
\author[0000-0003-4046-884X]{Harsha Blumer}
\affiliation{Department of Physics and Astronomy, West Virginia University, P.O. Box 6315, Morgantown, WV 26506, USA}
\affiliation{Center for Gravitational Waves and Cosmology, West Virginia University, Chestnut Ridge Research Building, Morgantown, WV 26505, USA}
\author{Adam Brazier}
\affiliation{Cornell Center for Astrophysics and Planetary Science and Department of Astronomy, Cornell University, Ithaca, NY 14853, USA}
\author[0000-0003-3053-6538]{Paul R. Brook}
\affiliation{Department of Physics and Astronomy, West Virginia University, P.O. Box 6315, Morgantown, WV 26506, USA}
\affiliation{Center for Gravitational Waves and Cosmology, West Virginia University, Chestnut Ridge Research Building, Morgantown, WV 26505, USA}
\author[0000-0003-4052-7838]{Sarah Burke-Spolaor}
\affiliation{Department of Physics and Astronomy, West Virginia University, P.O. Box 6315, Morgantown, WV 26506, USA}
\affiliation{Center for Gravitational Waves and Cosmology, West Virginia University, Chestnut Ridge Research Building, Morgantown, WV 26505, USA}
\author[0000-0003-0909-5563]{Bence Bécsy}
\affiliation{Department of Physics, Oregon State University, Corvallis, OR 97331, USA}
\author[0000-0002-5557-4007]{J. Andrew Casey-Clyde}
\affiliation{Department of Physics, University of Connecticut, 196 Auditorium Road, U-3046, Storrs, CT 06269-3046, USA}
\author[0000-0003-3579-2522]{Maria Charisi}
\affiliation{Department of Physics and Astronomy, Vanderbilt University, 2301 Vanderbilt Place, Nashville, TN 37235, USA}
\author[0000-0002-2878-1502]{Shami Chatterjee}
\affiliation{Cornell Center for Astrophysics and Planetary Science and Department of Astronomy, Cornell University, Ithaca, NY 14853, USA}
\author[0000-0002-3118-5963]{Siyuan Chen}
\affiliation{Kavli Institute for Astronomy and Astrophysics, Peking University, Beijing, 100871 China}
\author[0000-0002-4049-1882]{James M. Cordes}
\affiliation{Cornell Center for Astrophysics and Planetary Science and Department of Astronomy, Cornell University, Ithaca, NY 14853, USA}
\author[0000-0002-7435-0869]{Neil J. Cornish}
\affiliation{Department of Physics, Montana State University, Bozeman, MT 59717, USA}
\author[0000-0002-2578-0360]{Fronefield Crawford}
\affiliation{Department of Physics and Astronomy, Franklin \& Marshall College, P.O. Box 3003, Lancaster, PA 17604, USA}
\author[0000-0002-6039-692X]{H. Thankful Cromartie}
\affiliation{University of Virginia, Department of Astronomy, P.O. Box 400325, Charlottesville, VA 22904, USA}
\author[0000-0002-2185-1790]{Megan E. DeCesar}
\affiliation{George Mason University, Fairfax, VA 22030, resident at the Naval Research Laboratory, Washington, DC 20375, USA}
\author[0000-0002-6664-965X]{Paul B. Demorest}
\affiliation{National Radio Astronomy Observatory, 1003 Lopezville Rd., Socorro, NM 87801, USA}
\author[0000-0001-8885-6388]{Timothy Dolch}
\affiliation{Department of Physics, Hillsdale College, 33 E. College Street, Hillsdale, MI 49242, USA}
\affiliation{Eureka Scientific, 2452 Delmer Street, Suite 100, Oakland, CA 94602-3017, USA}
\author{Brendan Drachler}
\affiliation{School of Physics and Astronomy, Rochester Institute of Technology, Rochester, NY 14623, USA}
\affiliation{Laboratory for Multiwavelength Astrophysics, Rochester Institute of Technology, Rochester, NY 14623, USA}
\author{Justin A. Ellis}
\affiliation{Infinia ML, 202 Rigsbee Avenue, Durham NC, 27701}
\author[0000-0001-7828-7708,]{E. C. Ferrara}
\affiliation{Department of Astronomy, University of Maryland, College Park, MD, 20742, USA}
\affiliation{Center for Exploration and Space Studies (CRESST), NASA/GSFC, Greenbelt, MD 20771, USA}
\affiliation{NASA Goddard Space Flight Center, Greenbelt, MD 20771, USA}
\author[0000-0001-5645-5336]{William Fiore}
\affiliation{Department of Physics and Astronomy, West Virginia University, P.O. Box 6315, Morgantown, WV 26506, USA}
\affiliation{Center for Gravitational Waves and Cosmology, West Virginia University, Chestnut Ridge Research Building, Morgantown, WV 26505, USA}
\author[0000-0001-8384-5049]{Emmanuel Fonseca}
\affiliation{Department of Physics and Astronomy, West Virginia University, P.O. Box 6315, Morgantown, WV 26506, USA}
\affiliation{Center for Gravitational Waves and Cosmology, West Virginia University, Chestnut Ridge Research Building, Morgantown, WV 26505, USA}
\author[0000-0001-7624-4616]{Gabriel E. Freedman}
\affiliation{Center for Gravitation, Cosmology and Astrophysics, Department of Physics, University of Wisconsin-Milwaukee,\\ P.O. Box 413, Milwaukee, WI 53201, USA}
\author[0000-0001-6166-9646]{Nathan Garver-Daniels}
\affiliation{Department of Physics and Astronomy, West Virginia University, P.O. Box 6315, Morgantown, WV 26506, USA}
\affiliation{Center for Gravitational Waves and Cosmology, West Virginia University, Chestnut Ridge Research Building, Morgantown, WV 26505, USA}
\author[0000-0001-8158-638X]{Peter A. Gentile}
\affiliation{Department of Physics and Astronomy, West Virginia University, P.O. Box 6315, Morgantown, WV 26506, USA}
\affiliation{Center for Gravitational Waves and Cosmology, West Virginia University, Chestnut Ridge Research Building, Morgantown, WV 26505, USA}
\author[0000-0003-4090-9780]{Joseph Glaser}
\affiliation{Department of Physics and Astronomy, West Virginia University, P.O. Box 6315, Morgantown, WV 26506, USA}
\affiliation{Center for Gravitational Waves and Cosmology, West Virginia University, Chestnut Ridge Research Building, Morgantown, WV 26505, USA}
\author[0000-0003-1884-348X]{Deborah C. Good}
\affiliation{Department of Physics and Astronomy, University of British Columbia, 6224 Agricultural Road, Vancouver, BC V6T 1Z1, Canada}
\author[0000-0002-1146-0198]{Kayhan Gültekin}
\affiliation{University of Michigan, Dept. of Astronomy, 1085 S. University Ave., Ann Arbor, MI, 48104, USA}
\author[0000-0003-2742-3321]{Jeffrey S. Hazboun}
\affiliation{Department of Physics, Oregon State University, Corvallis, OR 97331, USA}
\author[0000-0003-1082-2342]{Ross J. Jennings}
\affiliation{Cornell Center for Astrophysics and Planetary Science and Department of Astronomy, Cornell University, Ithaca, NY 14853, USA}
\author[0000-0002-7445-8423]{Aaron D. Johnson}
\affiliation{Center for Gravitation, Cosmology and Astrophysics, Department of Physics, University of Wisconsin-Milwaukee,\\ P.O. Box 413, Milwaukee, WI 53201, USA}
\affiliation{Theoretical AstroPhysics Including Relativity (TAPIR), MC 350-17, California Institute of Technology, Pasadena, California 91125, USA}
\author[0000-0001-6607-3710]{Megan L. Jones}
\affiliation{Center for Gravitation, Cosmology and Astrophysics, Department of Physics, University of Wisconsin-Milwaukee,\\ P.O. Box 413, Milwaukee, WI 53201, USA}
\author[0000-0002-3654-980X]{Andrew R. Kaiser}
\affiliation{Department of Physics and Astronomy, West Virginia University, P.O. Box 6315, Morgantown, WV 26506, USA}
\affiliation{Center for Gravitational Waves and Cosmology, West Virginia University, Chestnut Ridge Research Building, Morgantown, WV 26505, USA}
\author[0000-0001-6295-2881]{David L. Kaplan}
\affiliation{Center for Gravitation, Cosmology and Astrophysics, Department of Physics, University of Wisconsin-Milwaukee,\\ P.O. Box 413, Milwaukee, WI 53201, USA}
\author[0000-0002-6625-6450]{Luke Zoltan Kelley}
\affiliation{Department of Astronomy, University of California at Berkeley, Berkeley, CA 94720, USA}
\affiliation{Center for Interdisciplinary Exploration and Research in Astrophysics (CIERA), Northwestern University, Evanston, IL 60208}
\author[0000-0003-0123-7600]{Joey Shapiro Key}
\affiliation{University of Washington Bothell, 18115 Campus Way NE, Bothell, WA 98011, USA}
\author[0000-0002-9197-7604]{Nima Laal}
\affiliation{Department of Physics, Oregon State University, Corvallis, OR 97331, USA}
\author[0000-0003-0721-651X]{Michael T. Lam}
\affiliation{School of Physics and Astronomy, Rochester Institute of Technology, Rochester, NY 14623, USA}
\affiliation{Laboratory for Multiwavelength Astrophysics, Rochester Institute of Technology, Rochester, NY 14623, USA}
\author[0000-0003-1096-4156]{William G Lamb}
\affiliation{Department of Physics and Astronomy, Vanderbilt University, 2301 Vanderbilt Place, Nashville, TN 37235, USA}
\author{T. Joseph W. Lazio}
\affiliation{Jet Propulsion Laboratory, California Institute of Technology, 4800 Oak Grove Drive, Pasadena, CA 91109, USA}
\affiliation{Theoretical AstroPhysics Including Relativity (TAPIR), MC 350-17, California Institute of Technology, Pasadena, California 91125, USA}
\author[0000-0003-0771-6581]{Natalia Lewandowska}
\affiliation{Department of Physics, State University of New York at Oswego, Oswego, NY, 13126, USA}
\author[0000-0001-5766-4287]{Tingting Liu}
\affiliation{Center for Gravitation, Cosmology and Astrophysics, Department of Physics, University of Wisconsin-Milwaukee,\\ P.O. Box 413, Milwaukee, WI 53201, USA}
\author[0000-0003-1301-966X]{Duncan R. Lorimer}
\affiliation{Department of Physics and Astronomy, West Virginia University, P.O. Box 6315, Morgantown, WV 26506, USA}
\affiliation{Center for Gravitational Waves and Cosmology, West Virginia University, Chestnut Ridge Research Building, Morgantown, WV 26505, USA}
\author{Jing Luo}
\altaffiliation{Author is deceased}
\affiliation{Department of Astronomy \& Astrophysics, University of Toronto, 50 Saint George Street, Toronto, ON M5S 3H4, Canada}
\author[0000-0001-5229-7430]{Ryan S. Lynch}
\affiliation{Green Bank Observatory, P.O. Box 2, Green Bank, WV 24944, USA}
\author[0000-0003-2285-0404]{Dustin R. Madison}
\affiliation{Department of Physics and Astronomy, West Virginia University, P.O. Box 6315, Morgantown, WV 26506, USA}
\affiliation{Center for Gravitational Waves and Cosmology, West Virginia University, Chestnut Ridge Research Building, Morgantown, WV 26505, USA}
\author[0000-0001-5481-7559]{Alexander McEwen}
\affiliation{Center for Gravitation, Cosmology and Astrophysics, Department of Physics, University of Wisconsin-Milwaukee,\\ P.O. Box 413, Milwaukee, WI 53201, USA}
\author[0000-0001-7697-7422]{Maura A. McLaughlin}
\affiliation{Department of Physics and Astronomy, West Virginia University, P.O. Box 6315, Morgantown, WV 26506, USA}
\affiliation{Center for Gravitational Waves and Cosmology, West Virginia University, Chestnut Ridge Research Building, Morgantown, WV 26505, USA}
\author[0000-0002-4307-1322]{Chiara M. F. Mingarelli}
\affiliation{Center for Computational Astrophysics, Flatiron Institute, 162 5th Avenue, New York, New York, 10010, USA}
\affiliation{Department of Physics, University of Connecticut, 196 Auditorium Road, U-3046, Storrs, CT 06269-3046, USA}
\author[0000-0002-3616-5160]{Cherry Ng}
\affiliation{Dunlap Institute for Astronomy and Astrophysics, University of Toronto, 50 St. George St., Toronto, ON M5S 3H4, Canada}
\author[0000-0002-6709-2566]{David J. Nice}
\affiliation{Department of Physics, Lafayette College, Easton, PA 18042, USA}
\author[0000-0002-4941-5333]{Stella Koch Ocker}
\affiliation{Cornell Center for Astrophysics and Planetary Science and Department of Astronomy, Cornell University, Ithaca, NY 14853, USA}
\author[0000-0002-2027-3714]{Ken D. Olum}
\affiliation{Institute of Cosmology, Department of Physics and Astronomy, Tufts University, Medford, MA 02155, USA}
\author[0000-0001-5465-2889]{Timothy T. Pennucci}
\affiliation{Institute of Physics, E\"{o}tv\"{o}s Lor\'{a}nd University, P\'{a}zm\'{a}ny P. s. 1/A, 1117 Budapest, Hungary}
\author[0000-0002-8826-1285]{Nihan S. Pol}
\affiliation{Department of Physics and Astronomy, Vanderbilt University, 2301 Vanderbilt Place, Nashville, TN 37235, USA}
\author[0000-0001-5799-9714]{Scott M. Ransom}
\affiliation{National Radio Astronomy Observatory, 520 Edgemont Road, Charlottesville, VA 22903, USA}
\author[0000-0002-5297-5278]{Paul S. Ray}
\affiliation{Space Science Division, Naval Research Laboratory, Washington, DC 20375-5352, USA}
\author{Joseph D. Romano}
\affiliation{Department of Physics and Astronomy, Texas Tech University, Lubbock, TX 79409-1051, USA}
\author[0000-0002-7283-1124]{Brent J. Shapiro-Albert}
\affiliation{Giant Army, 915A 17th Ave, Seattle WA 98122}
\affiliation{Department of Physics and Astronomy, West Virginia University, P.O. Box 6315, Morgantown, WV 26506, USA}
\affiliation{Center for Gravitational Waves and Cosmology, West Virginia University, Chestnut Ridge Research Building, Morgantown, WV 26505, USA}
\author[0000-0002-7778-2990]{Xavier Siemens}
\affiliation{Department of Physics, Oregon State University, Corvallis, OR 97331, USA}
\affiliation{Center for Gravitation, Cosmology and Astrophysics, Department of Physics, University of Wisconsin-Milwaukee,\\ P.O. Box 413, Milwaukee, WI 53201, USA}
\author[0000-0003-1407-6607]{Joseph Simon}
\affiliation{Jet Propulsion Laboratory, California Institute of Technology, 4800 Oak Grove Drive, Pasadena, CA 91109, USA}
\affiliation{Theoretical AstroPhysics Including Relativity (TAPIR), MC 350-17, California Institute of Technology, Pasadena, California 91125, USA}
\author[0000-0002-1530-9778]{Magdalena Siwek}
\affiliation{Center for Astrophysics, Harvard University, Cambridge, MA 02138, USA}
\author[0000-0002-6730-3298]{Ren\'{e}e Spiewak}
\affiliation{Jodrell Bank Centre for Astrophysics, Department of Physics and Astronomy, University of Manchester, Manchester M13 9PL, UK}
\author[0000-0001-9784-8670]{Ingrid H. Stairs}
\affiliation{Department of Physics and Astronomy, University of British Columbia, 6224 Agricultural Road, Vancouver, BC V6T 1Z1, Canada}
\author[0000-0002-1797-3277]{Daniel R. Stinebring}
\affiliation{Department of Physics and Astronomy, Oberlin College, Oberlin, OH 44074, USA}
\author[0000-0002-7261-594X]{Kevin Stovall}
\affiliation{National Radio Astronomy Observatory, 1003 Lopezville Rd., Socorro, NM 87801, USA}
\author[0000-0002-1075-3837]{Joseph K. Swiggum}
\altaffiliation{NANOGrav Physics Frontiers Center Postdoctoral Fellow}
\affiliation{Department of Physics, Lafayette College, Easton, PA 18042, USA}
\author[0000-0002-3360-9299]{Jessica Sydnor}
\affiliation{Department of Physics and Astronomy, West Virginia University, P.O. Box 6315, Morgantown, WV 26506, USA}
\affiliation{Center for Gravitational Waves and Cosmology, West Virginia University, Chestnut Ridge Research Building, Morgantown, WV 26505, USA}
\author[0000-0003-0264-1453]{Stephen R. Taylor}
\affiliation{Department of Physics and Astronomy, Vanderbilt University, 2301 Vanderbilt Place, Nashville, TN 37235, USA}
\author[0000-0002-2451-7288]{Jacob E. Turner}
\affiliation{Department of Physics and Astronomy, West Virginia University, P.O. Box 6315, Morgantown, WV 26506, USA}
\affiliation{Center for Gravitational Waves and Cosmology, West Virginia University, Chestnut Ridge Research Building, Morgantown, WV 26505, USA}
\author[0000-0002-4162-0033]{Michele Vallisneri}
\affiliation{Jet Propulsion Laboratory, California Institute of Technology, 4800 Oak Grove Drive, Pasadena, CA 91109, USA}
\affiliation{Theoretical AstroPhysics Including Relativity (TAPIR), MC 350-17, California Institute of Technology, Pasadena, California 91125, USA}
\author[0000-0003-4700-9072]{Sarah J. Vigeland}
\affiliation{Center for Gravitation, Cosmology and Astrophysics, Department of Physics, University of Wisconsin-Milwaukee,\\ P.O. Box 413, Milwaukee, WI 53201, USA}
\author[0000-0001-9678-0299]{Haley M. Wahl}
\affiliation{Department of Physics and Astronomy, West Virginia University, P.O. Box 6315, Morgantown, WV 26506, USA}
\affiliation{Center for Gravitational Waves and Cosmology, West Virginia University, Chestnut Ridge Research Building, Morgantown, WV 26505, USA}
\author[0000-0003-1551-1340]{Gregory Walsh}
\affiliation{Department of Physics and Astronomy, West Virginia University, P.O. Box 6315, Morgantown, WV 26506, USA}
\affiliation{Center for Gravitational Waves and Cosmology, West Virginia University, Chestnut Ridge Research Building, Morgantown, WV 26505, USA}

\author[0000-0002-6020-9274]{Caitlin A. Witt$^{\color{magenta}\S}$}
\affiliation{Center for Interdisciplinary Exploration and Research in Astrophysics (CIERA), Northwestern University, Evanston, IL 60208}
\affiliation{Adler Planetarium, 1300 S. DuSable Lake Shore Dr., Chicago, IL 60605, USA}
\affiliation{Department of Physics and Astronomy, West Virginia University, P.O. Box 6315, Morgantown, WV 26506, USA}
\affiliation{Center for Gravitational Waves and Cosmology, West Virginia University, Chestnut Ridge Research Building, Morgantown, WV 26505, USA}

\author[0000-0002-0883-0688]{Olivia Young}
\affiliation{School of Physics and Astronomy, Rochester Institute of Technology, Rochester, NY 14623, USA}
\affiliation{Laboratory for Multiwavelength Astrophysics, Rochester Institute of Technology, Rochester, NY 14623, USA}

\collaboration{1000}{The NANOGrav Collaboration}
%\altaffiliation{Author order alphabetical by surname}
\noaffiliation

% \todo{update with new full members}

\begin{abstract}
Pulsar timing array collaborations, such as the North American Nanohertz Observatory for Gravitational Waves (NANOGrav), are seeking \rev{to detect} nanohertz gravitational waves emitted by supermassive black hole binaries formed in the aftermath of galaxy mergers.
%are the only current observatories that can detect nanohertz gravitational waves. 
%Nanohertz gravitational waves emitted by supermassive black hole binaries can be detected by pulsar timing arrays such as the North American Nanohertz Observatory for Gravitational Waves (NANOGrav).
We have searched for continuous waves from individual circular supermassive black hole binaries using NANOGrav's recent 12.5-year data set. We created new methods to accurately model the uncertainties on pulsar distances in our analysis, and we implemented new techniques to account for a common red noise process in pulsar timing array data sets while searching for deterministic gravitational wave signals, including continuous waves. As we found no evidence for continuous waves in our data, we placed 95\% upper limits on the strain amplitude of continuous waves emitted by these sources. At our most sensitive frequency of 7.65 nanohertz, we placed a sky-averaged limit of $h_0 < $ \lowesth{}, and $h_0 <$ \bestpix{} in our most sensitive sky location. Finally, we placed a multi-messenger limit of \ultargetcrn{} on the chirp mass of the supermassive black hole binary candidate 3C~66B. 

\end{abstract}


\section{Introduction}
Supermassive black hole binaries (SMBHBs) are expected to form in the aftermath of galaxy mergers, when the two constituent supermassive black holes eventually become gravitationally bound \citep{Begelman1980}. If they are able to reach an advanced stage of evolution, with sub-parsec orbital separations, these binaries are predicted to be among the brightest sources of low-frequency gravitational waves (GWs) in the universe, emitting at frequencies of \rev{$\sim10^{-9}-10^{-7}$ Hz}. The GWs emitted by discrete SMBHBs are known as continuous waves (CWs) due to their minimal frequency evolution, while the dominant source of nanohertz GWs is expected to be the stochastic background of GWs (GWB) that has contributions from the entire cosmic population of SMBHBs and potentially other sources \citep{sesana_04,review}.

By carefully monitoring the radio pulses from stable millisecond pulsars (MSPs) over many years, pulsar timing arrays (PTAs) should be able to detect correlated fluctuations in the pulse times of arrival due to the influence of low-frequency GWs \citep{PTAs, ptas2}. There are multiple PTA collaborations currently operating; among them, the North American Nanohertz Observatory for Gravitational Waves \citep[NANOGrav;][]{nanograv}, the Parkes Pulsar Timing Array \citep[PPTA;][]{PPTA, ppta_2}, and the European Pulsar Timing Array \citep[EPTA;][]{EPTA} have each produced multiple pulsar timing data sets \rev{which have been searched for GWs.}
%with which to search for GWs. 
%These three groups along with the Indian Pulsar Timing Array \citep[InPTA;][]{inpta}, also combine efforts as a consortium known as the International Pulsar Timing Array 
These groups, along with other pulsar timing projects, combine efforts as a consortium known as the International Pulsar Timing Array \citep[IPTA;][]{ipta}. 

These PTA data sets have enabled numerous searches for GWs from SMBHBs, as well as primordial GWs \citep[e.g.][]{,vagnozzi,benetti}, cosmic strings \citep[e.g.,][]{11yr_gwb}, and cosmological phase transitions \citep{phase_transitions,ppta_phase_transitions}. %While there have been efforts to seek each of these signals in PTA data, the majority have focused on the GWB, as 
Modeling has suggested that the GWB signal from SMBHBs will be detected first \citep{rosado}. While PTAs have not yet detected a GWB, they have placed steadily improving limits on such a signal \citep{epta_gwb, 5yr_gwb, ppta_gwb_2013, epta_gwb_2015, shannon_2015, ipta_gwb_1, nano9year-GWB, 11yr_gwb} until around 2015, when published limits began to stabilize at a characteristic strain value of a few times $10^{-15}$. In the NANOGrav 12.5-year data set \citep{12p5_narrow}, PPTA second data release \citep{ppta_dr2}, EPTA data release 2 \citep{epta_crn}, and IPTA data release 2 \citep{dr2}, not only does the upper limit no longer decrease, but a common red noise (CRN) process with characteristics similar to those predicted for a SMBHB-origin GWB was detected to high significance, albeit without evidence for the specific spatial correlation assumed for the GWB \citep{12p5_gwb, ppta_dr2_gwb, dr2_gwb, ipta2cw}. \rev{Significant effort has been dedicated to determining if the CRN is an early warning sign of a future GWB detection \citep{astro4cast} or an anomaly due to pulsar noise modeling 
\citep{zic} and to the development of validity tests to determine between these two scenarios \citep{Goncharov}, but more data are required to make a final determination.}

While this common red-noise process is heartening for future GWB searches \citep{astro4cast}, it has sparked new challenges for CW searches, as the background takes the form of a noise process, which (like any noise process underlying a signal) will work to disrupt the sensitivity of CW searches. Over the past decades, all-sky and all-frequency CW searches have improved their sensitivity by several orders of magnitude in GW strain \citep[e.g.,][]{yardley+10,5yrCW, PPTA-cw-paper, EPTA-CW-paper, 11yrCW}, allowing the sensitivity horizon of PTAs to expand by several orders of magnitude. This has allowed the PTA horizon to include increasing numbers of specific systems of interest \citep[e.g.,][]{lommenbacker,Jenet2004,11yrCW,charisi_prep}.
\revtwo{PTAs are likely to reach the sensitivities required to detect a CW soon after the GWB is detected, 
{with recent studies suggesting this will occur in the next 5-10 years} \citep{rosado,mingarelli_2017,kelley,becsypop}. Additionally, we are working to revise and improve CW search methodologies
{through search speed-ups \citep{quickcw} and efficient sampling techniques \citep{11yrCW}} as CW upper limits decrease.}
%Even without a CW detection, informative astrophysical limits have been placed on the nearby population of SMBHBs \citep{11yrCW} as well as individual SMBHB candidates \citep{Jenet2004, 3c66b}.

In this paper, we present the results of an all-sky search for CWs from individual circular SMBHBs in the NANOGrav 12.5-year data set. This work is an extension of the searches performed in previous NANOGrav datasets (presented in \citealt{5yrCW} and \citealt{11yrCW} for the 5- and 11-year data sets, respectively), and uses analogous techniques to the search for CWs in the IPTA data release 2 \citep{ipta2cw}. Our new search benefited from the use of the more sensitive 12.5-year data set. Most critically, however, in this work we needed to account for the existence of an emerging common-noise signal in this data set, and understand the impact that this signal may have on CW sensitivity.
%, we seek to expand the methods used in previous searches to conduct a search for CWs that accounts for the emerging common signal in the data set.

This paper is organized as follows. In \autoref{sec:methods}, we present an overview of the data used for our analysis, details of new pulsar distance modeling methods created for CW searches, and a description of the GW signals and analysis methods used throughout this paper. In \autoref{sec:results}, we present the results of our GW searches, and in \autoref{sec:astro}, interpret their broader astrophysical context. For the busy reader, our main results can be summarized as follows:
\begin{itemize}
    \item For accurate low-frequency CW searches, the CRN that has been seen in GWB searches must be accounted for in our signal modeling; otherwise, our detection metrics may report a false positive result.
    \item Once the CRN was taken into account, we found that no CWs were detected in the 12.5-year data set.
    \item With this knowledge, we placed stringent limits on the CW amplitude as a function of GW frequency. For the most sensitive frequency of \bestfreq{}, we reach strain 95\% upper limits of $h_0<~$\lowesth{}, and we also placed limits on the CW amplitude at this frequency as a function of sky location. 
    \item While our all-sky sensitivity has improved with each subsequent NANOGrav data set, we found herein that for \revtwo{85\%} of the sky, the upper limit at the most sensitive frequency of $7.65\times 10^{-9}$ Hz is comparable to or worse than in previous data sets. 
    %Through extensive simulations, we linked this effect to the newly-detectable CRN process in the 12.5-year data set. \revtwo{
    %While the pulsar white noise has decreased compared to the 11-year data set, at this frequency the presence of the newly-detectable CRN process results in an increase of the UL across most of the sky, compared to the 11-year data set.}    
    Through extensive simulations \revtwo{
    %and despite complexities in comparing different PTA data sets due to non-identical pulsar noise properties
    that encompass the complex evolution of pulsar noise parameters, ephemeris updates, and Bayesian modeling}, we linked this effect to the newly-detectable CRN process in the 12.5-year data set.
    \item We used these limits to make inferences about the local population of SMBHBs, and limited the distance to an SMBHB emitting at $7.65\times 10^{-9}$ Hz to be greater than 86.7 Mpc for a $10^9 M_\odot$ binary in the most sensitive sky location.
    \item We used multi-messenger techniques to update limits on the chirp mass of the SMBHB candidate 3C~66B to be less than \ultargetcrn, and placed new limits on the chirp mass of SMBHB candidate HS 1630+2355 to be less than $\mathcal{M}<~(1.28 \pm 0.03) \times 10^{10} M_\odot. $
    % \item While our all-sky sensitivity has improved with each subsequent NANOGrav data set, we find herein that for a portion of the sky at low frequencies, there is a \emph{sensitivity noise floor} when comparing the previous NANOGrav data set to this NANOGrav data set. That is, the upper limits in about 3/4 of the sky (that which NANOGrav is most sensitive) do not improve from the previous data set. After investigation we believe that this noise floor is due to the existence of the common process in this data set. If this is true, the issue will not exist at higher frequencies, because the red-noise process will have a smaller amplitude. If the noise floor is caused by the GWB, we can liken the noise floor as something akin to low-frequency confusion noise. While our search technique did not aim to model and remove this noise floor, if it is sky-correlated (as a genuine GWB would be), it may be possible in the future to account for this signal and remove it.
\end{itemize}
In \autoref{sec:discussion}, we discuss the implications of these results.
In \autoref{sec:conclusion}, we summarize our conclusions.

\section{Methods}\label{sec:methods}

\subsection{The 12.5-year Data Set}

We analyzed the NANOGrav 12.5-year data set, originally published as \citet{12p5_narrow, 12p5_wide}, which consists of times-of-arrival (TOAs) and timing models from 47 pulsars. Two versions of the data set were created from the original observations, taken between 2004 and 2017, using independent analyses. Here, we make use of the narrowband version of the data set \citep{12p5_narrow}. This adds 2 pulsars and 1.5 years of observations over the previous 11-year data set. For GW analyses, we require the pulsars to have a timing baseline of at least 3 years; therefore, we use only 45 of the 47 pulsars included in the full data set. However, the 11-year data set included only 34 pulsars that could be used in GW analyses, so this addition, which includes a factor of $\sim1.5$ increase in the number of pulse TOAs, represents a significant addition of data, increasing our sensitivity.
It is important to note that the 12.5-year data set is not merely an addition of TOAs to previous releases, but a full re-analysis with an updated pipeline, described in detail in \citet{12p5_narrow}. Thus, our search also benefited from improved timing precision for pulsars shared with previous data sets. \rev{However, it is important to note that this reprocessing resulted in new values being measured for each pulsar's noise parameters.}
% Due to the upgraded processing methods, \todo{expect more sensitivity? Clearly need regular analyses? idk something something}

\subsection{Signal Model}

As in previous NANOGrav searches for continuous gravitational waves, we will describe the effect of an individual SMBHB on a pulsar's TOAs and its timing model.  A starting point is the residuals, $\delta t$, obtained after subtracting a basic timing model (which excludes noise and GW parameters) from the measured arrival times. While the methods remain nearly identical to previous iterations, slight alterations have been made to improve consistency with other work in the field, to reflect more recent data, and to include the CRN in the CW search. As such, we will lay out the methods with particular focus on any instances that have changed since NANOGrav's most recent CW search \citep{11yrCW}. Note that throughout this paper, we use units where $G = c = 1$, cosmology calculations assume \rev{$H_0 = 69.32 \mathrm{~km~s^{-1}~Mpc^{-1}}$}, and the GW derivations assume General Relativity.

The pulsar residuals can be separated into multiple components as 
\begin{equation}
    \delta t = M \epsilon + n_{\mathrm{white}} + n_{\mathrm{red}} + s,
\end{equation}
where $M$ is the design matrix, which describes the linearized timing model, and $\epsilon$ is a vector of the timing model parameter offsets. This term allows the timing model parameters of each pulsar to be adjusted in accordance with the presence of any additional signals. The variables $n_{\mathrm{white}}$ and $n_{\mathrm{red}}$ refer to vectors describing the pulsar white and red noise, respectively, and $s$ is a vector of GW-induced signal present in the residuals. 

\subsubsection{CW Signal}

For a GW source located at right ascension $\alpha$ and declination $\delta$, we define the polar angle $\theta = \pi/2 - \delta$ and azimuthal angle $\phi = \alpha$. The strain of GWs emitted from such a source can be written in terms of two polarizations as

\begin{equation}
    h_{a b}(t, \hat{\Omega})=e_{a b}^{+}(\hat{\Omega}) h_{+}(t, \hat{\Omega})+e_{a b}^{\times}(\hat{\Omega}) h_{\times}(t, \hat{\Omega}),
\end{equation}
where $\hat{\Omega}$ is a unit vector pointing from the the GW source to the Earth (along the direction of propagation), $h_{+,\times}$ are the polarization amplitudes, and $e_{a b}^{+, \times}$ are the polarization tensors. These can be written in the solar system barycenter frame as 
\begin{equation}
\begin{aligned} e_{a b}^{+} &=\hat{p}_{a} \hat{p}_{b}-\hat{q}_{a} \hat{q}_{b} \\ e_{a b}^{\times} &=\hat{p}_{a} \hat{q}_{b}+\hat{q}_{a} \hat{p}_{b}, \end{aligned}
\end{equation}
and are constructed from basis vectors. 
\begin{equation}
\begin{aligned} \hat{n}=&(\sin \theta \cos \phi, \sin \theta \sin \phi, \cos \theta) = -\hat{\Omega} \\ \hat{p}=&(\cos \psi \cos \theta \cos \phi-\sin \psi \sin \phi,\\ &\cos \psi \cos \theta \sin \phi+\sin \psi \cos \phi,-\cos \psi \sin \theta) \\ \hat{q}=&(\sin \psi \cos \theta \cos \phi+\cos \psi \sin \phi,\\ &\sin \psi \cos \theta \sin \phi-\cos \psi \cos \phi,-\sin \psi \sin \theta), \end{aligned}
\end{equation}
\rev{where $\psi$ is the GW polarization angle.} Note that this basis is different than that used in \citet{11yrCW} to maintain better consistency with previous references and the standards used by other GW detectors. Differences can be reduced to a rotation of the frame by an angle equivalent to the GW polarization angle $\psi$. These polarization tensors are used to construct the antenna pattern function $F^{+, \times}(\hat{\Omega})$, which describes the response of the pulsar (at unit vector $\hat{u}$) to the GW source, as in \citet{taylor2016}, where
\begin{equation}
F^{A}(\hat{\Omega}) \equiv \frac{1}{2} \frac{\hat{u}^{a} \hat{u}^{b}}{1+\hat{\Omega} \cdot \hat{u}} e_{a b}^{A}(\hat{\Omega}).
\end{equation}

Now, we can write the signal $s$ induced by the GW as seen in the pulsar's residuals as 
\begin{equation}
s(t, \hat{\Omega})=F^{+}(\hat{\Omega}) \Delta s_{+}(t)+F^{\times}(\hat{\Omega}) \Delta s_{\times}(t),
\end{equation}
where $\Delta s_{+, \times}$ is the difference between the signal induced at the Earth (the ``Earth term") and at the pulsar (the ``pulsar term"). This can be written as 
\begin{equation}
\Delta s_{+, \times}(t)=s_{+, \times}\left(t_{p}\right)-s_{+, \times}(t),
\end{equation}
where $t$ and $t_p$ represent the time when the GW passes the Earth and the pulsar, respectively. These times can be related geometrically by 
\begin{equation}
t_{p}=t-L(1+\hat{\Omega} \cdot \hat{u}),
\end{equation}
where $\hat{u}$ is the line of sight vector to the pulsar and $L$ is the distance to the pulsar (see \autoref{ss:distance} for further discussion of this value). 

For a circular binary at zeroth post-Newtonian (0-PN) order, $s_{+,\times}$ can be written as
\begin{equation}
\begin{aligned} s_{+}(t)=& \frac{\mathcal{M}^{5 / 3}}{d_{L} \omega(t)^{1 / 3}}\left[-\sin 2 \Phi(t)\left(1+\cos ^{2} \iota\right)\right], \\ s_{\times}(t)=& \frac{\mathcal{M}^{5 / 3}}{d_{L} \omega(t)^{1 / 3}}\left[2 \cos 2 \Phi(t) \cos \iota \right], \end{aligned}
\end{equation}
where $\iota$ is the inclination angle of the SMBHB, $d_L$ is the luminosity distance to the source, $\omega(t)$ and $\Phi(t)$ are the time-dependent angular orbital frequency and phase, respectively, and $\mathcal{M} \equiv\left(m_{1} m_{2}\right)^{3 / 5} /\left(m_{1}+m_{2}\right)^{1 / 5}$ is a combination of the two black hole masses known as the chirp mass. Again, note that the forms of these signals have been reorganized compared to those used in \citet{11yrCW}; due to the rotated frame of the antenna pattern functions now in use, they are equivalent. The variables $\mathcal{M}$ and $\omega$ refer to the redshifted values of these quantities, which relate to the rest-frame versions $\mathcal{M}_r$ and $\omega_r$ as 
\begin{equation}
    \begin{aligned} \mathcal{M}_{r} &=\frac{\mathcal{M}}{1+z}, \\ \omega_{r} &=\omega(1+z). \end{aligned}
\end{equation}
However, PTAs are currently \rev{thought to be sensitive only to} individual SMBHBs in the local universe where $(1+z)\sim 1$.

For a CW, the initial orbital angular $\omega_0$ frequency is related to the GW frequency by $\omega_0 = \pi f_{\rm GW}$, where $\omega_0 = \omega(t_0)$. For this search, we define the reference time $t_0$ as MJD 57933 (2017 June 29), \revtwo{the date of the last observation} for the 12.5-year data set. The time-dependent orbital phase and frequency of the binary are given by

\begin{equation}
    \begin{array}{l}\Phi(t)=\Phi_{0}+\frac{1}{32} \mathcal{M}^{-5 / 3}\left[\omega_{0}^{-5 / 3}-\omega(t)^{-5 / 3}\right], \\ \omega(t)=\omega_{0}\left(1-\frac{256}{5} \mathcal{M}^{5 / 3} \omega_{0}^{8 / 3} t\right)^{-3 / 8},\end{array}
\end{equation}
where $\Phi_0$ refers to the initial orbital phase \citep{5yrCW}. 
To account for the evolution of high chirp mass binaries over our observations, rather than assuming that there is no frequency evolution, we use the full expression for $\omega(t)$ as in \citet{11yrCW}.

\subsubsection{Noise Model}

For each individual pulsar, we model both white and red noise. We use a white noise model that is identical to that used in previous NANOGrav analyses, using three parameters: EFAC, EQUAD, and ECORR. EFAC scales the template-fitting TOA uncertainties induced by finite pulse signal-to-noise ratios by a multiplicative factor, EQUAD adds white noise in quadrature, and ECORR describes white noise that is correlated across TOAs derived from data collected simultaneously \citep{lam2017}.

For consistency with previous NANOGrav analyses, to model individual pulsar red noise, the noise spectrum is divided into 30 linearly spaced bins, ranging from $1/T_{\mathrm{obs}}$ to $30/T_{\mathrm{obs}}$, where $T_{\mathrm{obs}}$ is the total observation baseline for each pulsar. Then, the power spectral density of the red noise is fit to a power-law model as in \citet{shannon2010, lam2017}, where
\begin{equation}\label{eq:rn}
    P(f)=\frac{A_{\mathrm{red}}^{2}}{12\pi^2}\left(\frac{f}{f_{\mathrm{yr}}}\right)^{-\gamma_{\mathrm{red}}}~\mathrm{yr}^3.
\end{equation}
Here, $f_{\mathrm{yr}}\equiv 1/(1~\mathrm{year})$, $A_{\mathrm{red}}$ is the red noise amplitude, and $\gamma_{\mathrm{red}}$ is the power law spectral index. The prior on $A_{\mathrm{red}}$ is log-uniform in the range $[-20,-11]$, while the prior on $\gamma$ is uniform in the range [0,7].
%, with added jump distributions described in \citet{11yrCW}.

As mentioned above, for the first time, a CRN signal is now detectable in the 12.5-year data set \citep{12p5_gwb}. Because of this, we included a CRN term in our signal model for a portion of our analyses, \revtwo{where the CRN amplitude and spectral index are held fixed to those preferred in \citet{12p5_gwb}}. The results of searches that only model a CW necessitated this addition, and are described in detail in \autoref{sec:results}.
The power spectral density of the CRN 
\begin{equation}
    P(f)=\frac{A_{\mathrm{CRN}}^{2}}{12\pi^2}\left(\frac{f}{f_{\mathrm{yr}}}\right)^{-\gamma_{\mathrm{CRN}}}~\mathrm{yr}^3,
\end{equation}
takes the same form as that of the pulsar red noise in \autoref{eq:rn}, but with an amplitude $A_{\mathrm{CRN}}$ and spectral index $\gamma_{\mathrm{CRN}}$ that are common to all of the pulsars in the array.
\subsection{Bayesian Methods}
We utilized Bayesian inference techniques to determine the posterior distributions of GW parameters.
In previous CW analyses \citep{5yrCW, 11yrCW}, these results were compared to a frequentist metric, the $\mathcal{F}_p$ statistic \citep{ellis2012} to confirm our key results. However, as this method does not currently account for a common process other than a CW in the data, more development will be necessary to produce reliable frequentist results on the 12.5-year data set \rev{through the addition of noise-marginalization capabilities, similar to \citet{nmos}.}. Therefore, in this work, we will focus solely on the Bayesian searches, and the frequentist analyses will be presented in a future work.

In each analysis, we include the \textsc{BayesEphem} model \citep{bayesephem} to account for the uncertainties in the Solar System ephemeris, which, as first described in \citet{11yr_gwb}, can have large impacts on the computation of GW upper limits with PTAs. We used DE438 \citep{de438} plus \textsc{BayesEphem} to transform from individual observatory reference frames to an inertial frame centered at the Solar System Barycenter.

As in previous NANOGrav CW searches, we use the \texttt{enterprise} \citep{enterprise} package to construct the priors and evaluate the likelihood, which takes the same form as in \citet{11yrCW} and \citet{5yrCW}. The Markov Chain Monte Carlo (MCMC) sampler package \texttt{PTMCMCSampler} \citep{ptmcmc} was used to explore the parameter space. \rev{Before analyzing our data, we performed a prior-recovery analysis to ensure that the sampler could search the entire prior volume.}

The CW signal model can be described by nine global parameters:

\begin{equation}
    \{ \theta,\phi,f_{\rm GW},\Phi_0,\psi,i,\mathcal{M},d_{\rm L},{h_0}\},
\end{equation}

which describe the circular SMBHB's:

\begin{itemize}[noitemsep, nolistsep]
    \item position on the sky $(\theta, \phi)$;
    \item GW frequency, related to the orbital frequency at \rev{a} reference time $(f_{\rm GW})$;
    \item orbital phase at \rev{a} reference time $(\Phi_0)$;
    \item GW polarization angle $(\psi)$;
    \item orbital inclination $(\iota)$;
    \item chirp mass ($\mathcal{M}$);
    \item luminosity distance ($d_{\rm L}$);
    \item strain amplitude $(h_0)$, which is related to the chirp mass, GW frequency, and luminosity distance.
\end{itemize}

Since $h_0$ can be defined as 
\begin{equation}\label{eq:strain}
    h_0 = \frac{2 {\mathcal{M}^{5/3}}(\pi f_{{GW}})^{2/3}}{{d_L}}, 
\end{equation}
there is a degeneracy between $h_0$, $\mathcal{M}$, $f_{\rm GW}$, and $d_L$, and therefore
only eight of these parameters are required to fully describe the global CW signal. The following types of searches use a variety of prior setups to sample the necessary eight global parameters, and are described below and summarized in \autoref{tab:cw_priors}. \rev{Including the necessary parameters to model the red noise in and distance to each of the 45 pulsars and model uncertainties in the solar system ephemeris with \texttt{BayesEphem}, there are 198 parameters in our all-sky MCMCs.}

As in \citet{11yrCW}, to determine if a CW has been detected by any of our analyses, we first performed a detection analysis with the priors described in \autoref{tab:cw_priors}, with the key difference between this and upper limit analyses being a log-uniform prior on the strain amplitude of the CW. Then, we calculated the Bayes factor using the Savage-Dickey formula \citep{dickey}, 
\begin{equation}
    \mathcal{B}_{10} \equiv \frac{\text { evidence }\left[\mathcal{H}_{1}\right]}{\text { evidence }\left[\mathcal{H}_{0}\right]}=\frac{p\left(h_{0}=0 \mid \mathcal{H}_{1}\right)}{p\left(h_{0}=0 \mid \mathcal{D}, \mathcal{H}_{1}\right)}.
\end{equation}
Here, $\mathcal{H}_{1}$ is the model with a CW, $\mathcal{H}_{0}$ is the model without one,  $p\left(h_{0}=0 \mid \mathcal{H}_{1}\right)$ is the prior at $h_0=0$, and 
$p\left(h_{0}=0 \mid \mathcal{D}, \mathcal{H}_{1}\right)$ is the posterior at $h_0=0$. Since $\mathcal{H}_{1}$ and $\mathcal{H}_{0}$ are nested models (i.e., $\mathcal{H}_{0}$ is $\mathcal{H}_{1}: h_0=0$), we used the Savage-Dickey formula to estimate $p\left(h_{0}=0 \mid \mathcal{D}, \mathcal{H}_{1}\right)$ as the average fraction of samples in the lowest-amplitude bin in a histogram of $h_0$ samples for a range of bin sizes. We then computed the one-sigma error on the Bayes factor as 
\begin{equation}
    \sigma=\frac{\mathcal{B}_{10}}{\sqrt{n}},
\end{equation}
where $n$ is the number of samples in the lowest-amplitude bin. As with the Bayes factor values, the average error is computed for a range of histogram bin sizes. 

Throughout this work, we computed 95\% upper limits as the 95th percentile of relevant strain (or chirp mass, for multi-messenger analyses) posterior distributions. For these analyses, a uniform prior on the strain amplitude is used, which translates to a linear-exponential (LinExp) prior on $\mathrm{log_{10}}h$. The error on the 95\% upper limit, due to the finite number of samples, is calculated as 
\begin{equation}
    \sigma_{\mathrm{UL}}=\frac{\sqrt{x(1-x) / N_{s}}}{p\left(h_{0}=h_{0}^{95 \%} \mid \mathcal{D}\right)},
\end{equation}
where $x = 0.95$ and $N_s$ is the number of effective samples in the MCMC chain.

\subsubsection{All-Sky Searches}\label{methods:sky-avg}

To search for GWs from SMBHBs located in any direction, we use uniform priors on the source sky position $(\cos\theta, \phi)$, as well as the cosine of the source inclination $\cos\iota$, polarization angle $\psi$, and GW phase $\Phi_0$. We used log-uniform priors on $h_0$ for detection analyses, and uniform priors on $h_0$ for upper limit analyses, so as to set the most conservative upper limit. For both analysis types, priors on $\mathrm{log}_{10}(h_0)$ span the range $[-18,-11]$, which accounts for an over-conservative range around the sensitivity of the most recent data sets (order $-15$), and the minimum of which is well below our sensitivity.
%effectively a strain amplitude of 0 compared to our sensitivity.

We performed many searches at fixed values of $f_{\rm GW}$, to evaluate detection statistics and our sensitivity across the entire nanohertz GW band. The lowest frequency value was set by the time span of our data set, $f_{\rm GW} = 1/(12.9~\mathrm{years}) = 2.45 \times 10^{-9}$ Hz. The highest frequency value is limited by the observation cadence of our data (approximately one observation per 2--4 weeks). However SMBHBs at that frequency, at the mass range where their strains would be large enough to be detectable by PTAs, have exceedingly short inspiral timescales (a few weeks up to $\sim 3$ months). Thus, they are unlikely to be detectable in our data set \rev{due to rapid evolution, and therefore low residence times, at these frequencies, coupled with decreasing sensitivity of the PTA with increasing GW frequency caused by white noise} \citep{islo19, 11yrbwm}.
%will likely have either merged or be evolving extremely quickly, so they are unlikely to be detectable. 
Therefore, we set our maximum frequency to $3.178\times 10^{-7}$ Hz (equivalent to one GW cycle every $\sim36$ days and a GW inspiral time of $\sim34$ days). This is the same high-frequency cutoff value used in \citet{5yrCW, 11yrCW}.

For most of the frequency band, we searched over $\mathrm{log}_{10}({\mathcal{M}}/{M_\odot})$ with a log-uniform prior with a range of $[7,10]$. However, for very high-frequency sources, we limit the maximum value of the prior to account for high-chirp-mass binaries never emitting GWs at the highest frequencies in our band, as they will have merged prior to emitting GWs at the searched frequency. This cutoff is relevant at \rev{$f_{\rm GW}\geq 1.913
\times 10^{-7}$} Hz. Assuming binaries merge when the orbital frequency is equal to the innermost stable circular orbit (ISCO) frequency, $\mathcal{M}$ must satisfy
\begin{equation}
    \mathcal{M}_{\mathrm{max}} \leq \frac{1}{6^{3 / 2} \pi f_{\mathrm{GW}}}\left[\frac{q}{(1+q)^{2}}\right]^{3 / 5},
\end{equation}
where $q$ is the SMBHB mass ratio. Here, we calculated the chirp mass cutoff for $q=1$. 
%This adjustment to the $\mathcal{M}$ prior only affected searches where $f_{\rm GW}\geq 191.3$ nHz. 

\subsubsection{Sky Map }
Due to the non-uniform distribution of pulsars on the sky, the NANOGrav PTA is not equally sensitive in all directions. To analyze the differences in sensitivity, once detection analyses were completed, we placed upper limits on 768 pixels distributed isotropically across the sky using \texttt{healpy} \citep{healpix,healpy}; each pixel covers an area of 53.72 square degrees.
\rev{This value is chosen to balance \texttt{healpy}'s requirements for map transformations with our desired resolution. These settings allow us to resolve details on the sky sensitivity map, but not overwhelm our computational capabilities or explore much beyond the expected localization capabilities of PTAs \citep{sesanavecchio}.}
% \todo{describe healpy here or nah?}. 
We allowed the sampler to search a uniform prior across each of the 768 pixels, so as to still sample the entire sky across the entire analysis. 

Due to the large computational cost required to conduct 768 independent runs, the sky map is created at only a single frequency, and only upper limits are computed. We selected $7.65 \times 10^{-9}$ Hz, as it was the most sensitive in the sky-averaged analysis. As this is in the low-frequency regime where we expect the inclusion of the CRN to be significant, it is included in our signal model. All other modeling is done identically to \autoref{methods:sky-avg}, and is summarized in \autoref{tab:cw_priors}.

%!H

\subsubsection{Targeted Search}\label{sec:target_methods}

%Additionally, we performed a new search for the electromagnetic SMBHB candidate HS 1630+2355. First identified as a periodic quasar in \citep{Graham2015}, this candidate is identified as a top PTA CW candidate in \citet{xin_CRTS}

In addition to the two variations of searches described above, we also perform a targeted search for two known SMBHB candidates, 3C~66B and HS~1630+2355. Rather than a search for a generic SMBHB within a nearby galaxy cluster, as was done in \citet{11yrCW} and \citet{nanograv_nearby}, here, we targeted these binary candidates directly. 3C~66B was the subject of \citet{3c66b}, and was first identified because of observed orbital motion in the AGN core \citep{Sudou2003}. Here, we were able to provide an updated analysis with the addition of new data included in the 12.5-year data set. HS 1630+2355. was first identified as a periodic quasar in \citet{Graham2015}, and was identified as a top PTA CW candidate in \citet{xin_CRTS} due to \rev{its} location near our best-timed pulsars.

For the targeted search, we perform detection and upper limit analyses in the same way as in \autoref{methods:sky-avg}, with a few differences in the model priors. Because we know the sky location and luminosity distance to 3C~66B, as well as a frequency estimate, these parameters are set to constants in this search. This allows us to place constraints directly on the (observer-frame) chirp mass of the binary, rather than its GW strain amplitude. For a detection analysis, the prior on $\mathrm{log}_{10}\big({\mathcal{M}}/{M_\odot}\big)$ is log-uniform in the range $[7,10]$, while for upper limit analyses, the prior is uniform over this range. The remaining priors are identical to the above analyses, and are summarized in \autoref{tab:cw_priors}.

\begin{table*}
    \centering
    \begin{tabular}{c|ccccc}
    \hline
        &\multicolumn{2}{c}{All-Sky} & Sky Map & \multicolumn{2}{c}{Targeted}  \\
    \hline
    \hline
    
    Analysis Type & Detection & Upper Limit & Upper Limit & Detection & Upper Limit \\
    \hline
    CRN & Y/N & Y/N & Y & Y/N & Y/N \\
     $\mathrm{log_{10}}h$ & Uniform(--18,--11) & LinExp(--18,--11) & LinExp(--18,--11) & -- & -- \\
     $\mathrm{log_{10}}\mathcal{M}$ & Uniform(7,$\mathcal{M}_{\mathrm{max}}$) & Uniform(7,$\mathcal{M}_{\mathrm{max}}$) &
    Uniform(7,$\mathcal{M}_{\mathrm{max}}$) & 
    Uniform(7,$\mathcal{M}_{\mathrm{max}}$) & 
    LinExp(7,$\mathcal{M}_{\mathrm{max}}$) \\
     $\mathrm{log_{10}}d_L$ & -- & -- & -- & Constant & Constant\\
     $\mathrm{log_{10}}f_{\rm GW}$ & Constant (many) & Constant (many) & Constant (single) & Constant & Constant\\
     $\phi$ & Uniform(0,$2\pi$) & Uniform(0,$2\pi$) & Uniform(pixel)& Constant & Constant \\
     $\cos\theta$ & Uniform(--1,1) & Uniform(--1,1) & Uniform(pixel)& Constant & Constant \\
     $\psi$ & Uniform(0,$\pi$) & Uniform(0,$\pi$) & Uniform(0,$\pi$)& Uniform(0,$\pi$) & Uniform(0,$\pi$) \\
    $\Phi_0$ & Uniform(0,$2\pi$) & Uniform(0,$2\pi$) & Uniform(0,$2\pi$)& Uniform(0,$2\pi$) & Uniform(0,$2\pi$) \\
    $\cos\iota$ & Uniform(--1,1) & Uniform(--1,1) & Uniform(--1,1)& Uniform(--1,1) & Uniform(--1,1) \\
    \hline
    \end{tabular}
    \caption{CW parameter priors for each analysis.}
    \label{tab:cw_priors}
\end{table*}

\subsubsection{Pulsar Distance Priors}\label{ss:distance}

In this work, we adopted a data-driven approach to handle the large uncertainties on pulsar distance measurements, which, in addition to a phase at each pulsar, affect the modeling of the pulsar terms of the CW signal. As in previous searches, the pulsar distance was used as a free parameter in the search. This allowed us to marginalize over the pulsar distance, and avoid incorrect modeling of the signal at the the location of the pulsar. \rev{Without such modeling, we would be required to dramatically increase our prior volume to allow the pulsar distances to vary across galactic scales, or, if the incorrect value were assumed, risk losing our chance at detecting the pulsar term of the CW signal. This is particularly critical at high GW frequencies, where sources evolve significantly between the Earth and pulsar terms.}
%In certain types of analyses, such as the multi-messenger targeted searches for individual SMBHB candidates described in \autoref{sec:target_methods}, the pulsar term has been shown to be non-critical for a CW detection \citep{maria_prep}; }

In previous versions of this search \citep[e.g.][]{11yrCW, 3c66b}, the pulsar distance prior was constructed from a Gaussian scaled to the parallax distance and associated uncertainty listed in \citet{dist}; if no distance was listed, a value of $1.0\pm0.2$ kpc was assumed. While this assumption is reasonable while placing upper limits (see discussion within \citealt{3c66b}), as the PTA reaches sensitivities where a detection is nearly possible, an improvement was needed. 

In this work, every pulsar distance prior was constructed from a measurement or estimate. If a pulsar had a significant independent parallax measurement\footnote{\url{http://hosting.astro.cornell.edu/research/parallax/}, with values compiled from \citet{px1, px2, px3, px4, px5, px6, px7, px8, px9, px10, px11, px12, px13, px14, px15, px16, px17}}, such as from Very Long Baseline Interferometry (VLBI), or timing parallax measured in the 12.5-year data set, this value was used to construct a prior on pulsar distance ($L$).
\begin{equation}
p(L) = \frac{1}{\sqrt{2 \pi} \sigma_{\varpi} L^2} \exp\left[{\frac{-(PX-L^{-1})^2}{2 \sigma_{\varpi}^2}}\right], 
\end{equation}
which inverts the approximately Gaussian shape of a parallax prior to describe the prior for distance \citep{vigeland_pdist}. Here, significance was defined by the parallax measurement ($\varpi$) having an associated uncertainty ($\sigma_{\varpi})$ of less than 30\%, so as to avoid the introduction of any errors due to the Lutz-Kelker bias \citep{lk}. If multiple measurements of sufficient quality existed, these values and uncertainties were combined with a weighted average before being used to construct the parallax-distance prior, which ensures that the highest-quality measurements contribute the most to the resulting prior. 

If there are no parallax measurements that could be used to calculate the pulsar's distance, the pulsar's dispersion measure (DM) was used to construct a distance estimate using NE2001 \citep{ne2001} and, subsequently, the distance prior. Since these values are only an estimate, we constructed a broad, nearly uniform prior for the DM-distance value and a 20\% uncertainty \citep{ne2001, jones_dm, lam_dm}, with the shape
\begin{equation}
    p(L) =
\left\{
	\begin{array}{ll}
		\mathrm{Half-Gaussian}  & \mbox{if } L < 0.8~L_\mathrm{DM} \\
		\mathrm{Uniform}  & \mbox{if } 0.8~L_{DM} \leq L \leq 1.2~L_\mathrm{DM} \\
		\mathrm{Half-Gaussian} & \mbox{if } L > 1.2~L_\mathrm{DM}
	\end{array}
\right.
\end{equation}

% \begin{equation}
%     p(L) =
% \left\{
% 	\begin{array}{ll}
% 		a e^{\frac{L-0.8 d_{\rm DM}}{2b}}  & \mbox{if } L < 0.8~d_\mathrm{DM} \\
% 		a  & \mbox{if } 0.8~d_{DM} \leq L \leq 1.2~d_\mathrm{DM} \\
% 		a e^{\frac{L-1.2 d_{\rm DM}}{2b}} & \mbox{if } L > 1.2~d_\mathrm{DM}
% 	\end{array}
% \right.
% \end{equation}
Here, the Half-Gaussian additions have standard deviations of
%Here, $b$ is 
%$0.25\times$ 
\rev{one quarter of }the DM-distance uncertainty. Unlike a sharp boundary, these additions allowed the sampler to move into the edges of this prior range, which accounted for any differences in distance estimates by alternative electron density models, such as \citet{ymw}. While pulsar distance priors will still only induce minor influences on the results of an upper limit analysis \citep{3c66b}, by constructing new priors to accurately handle pulsar distance measurements and estimates, we have prepared our methods for a future detection of a CW, which will be more reliant on the pulsar term of the signal than upper limit evaluations. These values and the priors used are compiled in \autoref{tab:dists}.

% \subsection{Frequentist Statistic}
% As in \citet{5yrCW} and \citet{11yrCW}, we performed a frequentist analysis with the $\Fpstat$ to augment our Bayesian results. The $\Fpstat$, also computed with the software package \texttt{enterprise}, is an incoherent GW detection statistic that is derived by maximizing the log of the likelihood ratio \citep{ellis2012}. This method assumes that the SMBHB has no significant evolution over the timescale of our observations. 

% \update{
% In the absence of a signal, 2$\Fp$ is predicted to follow a \todo{non-central?} chi-square distribution with $2N_p$ degrees of freedom, where $N_p$ is the number of pulsars in the array. For the 12.5-year data set, $N_p$ is 45. The corresponding false-alarm probability (FAP) is 
% \begin{equation}
%     P_{F}\left(\mathcal{F}_{p, 0}\right)=\exp \left(-\mathcal{F}_{p, 0}\right) \sum_{k=0}^{N_{p}-1} \frac{\mathcal{F}_{p, 0}^{k}}{k !}.
% \end{equation}
% As we perform GW searches over the entire frequency range $2.45-315$ nHz at $N_f=129$ independent frequencies spaced by $1/T_{obs}$, we compute the $\Fpstat$ $N_f$ times. The FAP for the entire search is therefore
% \begin{equation}
%     P_{F}^{T}\left(\mathcal{F}_{p, 0}\right)=1-\left[1-P_{F}\left(\mathcal{F}_{p, 0}\right)\right]^{N_{f}}.
% \end{equation}
% }

\section{Results}\label{sec:results}

% Still need to do:
% \begin{enumerate}
%     \item Fstat
% \end{enumerate}

% Need to update final:
% \begin{enumerate}
%     \item BFs
%     \item ULs
%     \item 3c66b UL/BF
%     \item Most sensitive vs sky avg
% \end{enumerate}

\subsection{All-Sky Searches}\label{ss:all_sky}

\begin{figure}
    \centering
    \includegraphics[width = 1\columnwidth]{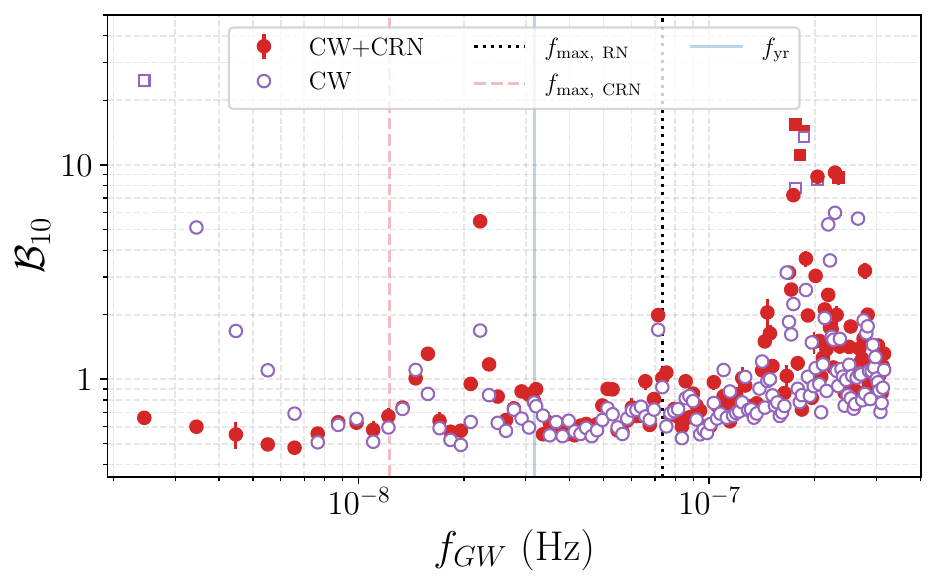}
    \caption{Savage-Dickey Bayes factors for a CW at each GW frequency. At low frequencies, inclusion of a CRN in the model (red) is necessary to avoid a false CW detection as in the CW-only model (unfilled purple). Square markers indicate a frequency where the initial analysis returned an undefined Savage-Dickey Bayes Factor, meaning the zoom-in analysis was necessary to calculate an accurate Bayes factor. With these methods, we found that no CWs are detected in the 12.5-year data set. \rev{Key frequencies are marked by vertical lines, including $f_\mathrm{yr}$ (blue solid), $f_\mathrm{max,~RN} = 30/T_{obs}$ (black dotted), and $f_\mathrm{max,~CRN} = 5/T_{obs}$ (red dashed).}} 
    \label{fig:bf}
\end{figure}

For each GW frequency in our search, we performed a detection analysis on the 12.5-year data which marginalized over the source sky location.  \autoref{fig:bf} shows the Bayes factor for a CW at each searched GW frequency in purple. It is important to note \rev{that the Bayes factor for $f_{\rm GW}= 2.45 \times 10^{-9}$ Hz (the lowest frequency analyzed) was undefined}, with a steady decrease in the following four frequency bins. Ordinarily, \rev{a very large (or undefined) Bayes factor} would be a first indication for the detection of a CW. However, given the strong evidence for the existence of a CRN process in the 12.5-year data set \citep{12p5_gwb}, it is clear that this signal appears to be of similar form; that is, what we have detected is bright at low frequencies and declines toward higher frequency. Once a common red-noise process is added to the model, with the $\mathrm{log}_{10}A_\mathrm{CRN}$ and $\gamma_{\mathrm{CRN}}$ parameters fixed to the maximum likelihood values ($-15.80$ and $6.08$, respectively) found by a search analogous to \citet{12p5_gwb}, the Bayes factors for a CW at low $f_{\rm GW}$ return to $<1$ (leftmost red points in the figure). Therefore, throughout this paper, we will present the results of many analyses with a fixed CRN included in our model. \rev{To constrain the \revtwo{initially undefined} Bayes factor at $f_{\rm GW} = 2.45$ nHz, we adapt the methodology described in \citet{zoom} to use a second MCMC analysis to ``zoom in" on the low end of the strain prior range by limiting the prior to the 10th percentile of the original posterior. Therefore, the posterior height at $h_0 = 0$ becomes
\begin{equation}
    p\left(h_{0}=0 \mid \mathcal{D}, \mathcal{H}_{1}\right)=\frac{n_{2}}{N_{2}} \frac{n_{1}}{N_{1}} \frac{1}{d h }, 
\end{equation}
with fractional uncertainty 
\begin{equation}
    \sqrt{\frac{1}{n_{1}}+\frac{1}{n_{2}}},
\end{equation}
where $N_1$ is the number of samples in the initial run and $n_1$ is the number of samples in the focused region (defined as the 10th percentile of the initial run). Then, $N_2$ is the number of samples in the focused run, with $n_2$ of those samples located in the lowest-amplitude bin of width $dh$.
}

% \newcommand{\highestBF}[1]{$\infty \pm \infty$}
% \newcommand{\highestBFfreq}[1]{$XXX$ nHz}

% Besides this instance, the highest Bayes factor was at $f_{\rm GW} = $ \highestBFfreq, with $\mathcal{B}_{10} = $ \highestBF, which is not enough to claim a detection. \todo{update after runs finish} 

%for thesis circulation; update after runs finish

We note that a few frequencies above $f_\mathrm{GW} = 1\times 10^{-7}$ Hz have $\mathcal{B}_{10}$ values that are returned as undefined. However, upon inspection, this is due to poor sampling in a few frequency bins, where the sampler does not explore low strain values, rather than a detection of a CW. This occurs in areas of parameter space where the likelihood is particularly complex and difficult to explore in a finite run-time
due to the numerous \revtwo{complexities} at $f_\mathrm{GW}>1\times 10^{-7}$ Hz, such as covariances between the CW likelihood with pulsar binary orbits and potential unmodeled red noise above the 30-frequency power law cutoff \citep{epta_noise}
Therefore, a few elevated Bayes factors are not unexpected. 
% With further sampling, it is expected that these values will settle near $\mathcal{B}_{10}\sim10$, and therefore, these results should be treated as preliminary for frequencies greater than $1\times 10^{-7}$nHz.
\rev{To mitigate this effect, we again apply the ``zoom-in" methodology described above. After this procedure, all frequencies have Bayes factor values of $\mathcal{B}_{10}\lesssim10$.}

The only frequency that needed this treatment for both the CW and CW+CRN models is $1.763\times10^{-7}$ Hz, which resulted in a Bayes factor of 15.43 in the CW+CRN case, and 7.79 in the CW-only case. While we inspected our analyses at this frequency with extra care, these Bayes factors are still relatively low compared to those required to claim a detection, especially since binaries at these high frequencies are expected to be quite rare \citep{kelley, becsypop}. For comparison, evidence in favor of a given model is generally not considered strong for Bayes factors $\lesssim 100$ \citep{kass}.
%To confirm, we analyzed this frequency with an identical signal model in NANOGrav's preliminary 15-year data set, and do not observe growing evidence for the same signal as we would expect if this Bayes factor were indicating the presence of a CW. 
Therefore, we will monitor this frequency in future data sets, but currently, our analyses indicate that no CWs are detected in the 12.5-year data set.
%acceptable mention of 15yr?

\begin{figure*}
    \centering
    \includegraphics[width = 1.5\columnwidth]{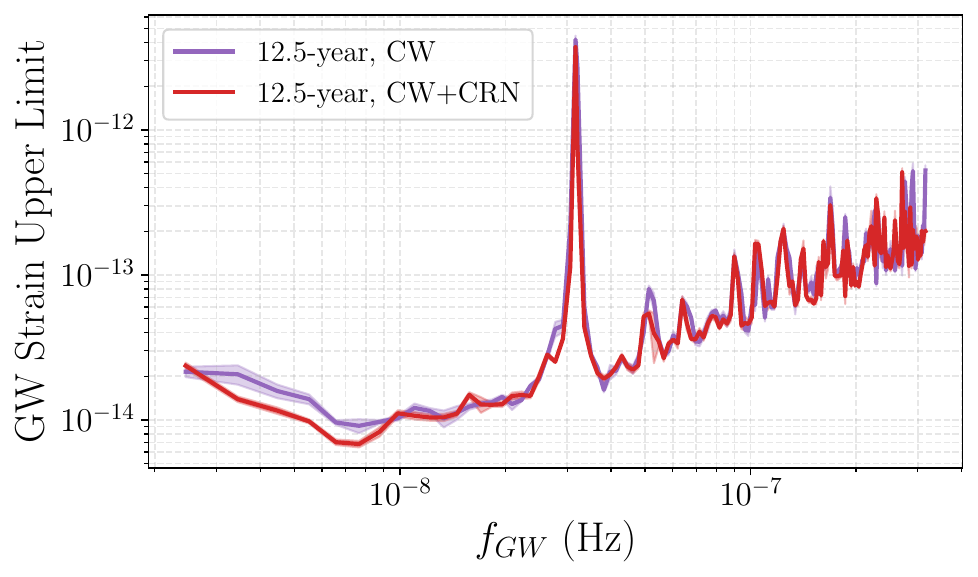}
    \caption{All-sky CW strain 95\% upper limits and associated error regions, with (red) and without (purple) a CRN included in the model. At low frequencies, modeling the CRN is necessary to avoid over-estimating our strain upper limits. We are the least sensitive to CWs at $f_\mathrm{GW}=$1/(1 year) due to the Earth's orbit, creating the large feature seen in this and other figures.}
    \label{fig:ul}
\end{figure*}

As we found no strong evidence for a GW from an individual SMBHB in the 12.5-year data set, we proceeded to place all-sky upper limits on the GW strain, with results shown in \autoref{fig:ul}. We again conduct this analysis using two different models, one which includes only a CW (purple) and one which includes both a CW and a CRN process (red). While in both cases, the most sensitive frequency (that with the lowest strain upper limit) is \bestfreq{}, the strain upper limits are lower when the CRN is included in the model. In this case, we can limit the strain to $h_0 <$ \lowesth{}, while when the CRN is neglected, the best limit we can place on CW strain is $h_0 <$ \lowesthonly{}. This trend of the CW+CRN model resulting in lower upper limits than a CW-only model continues until frequencies of approximately $1\times 10^{-8}$ Hz, above which, where the effect of the power-law CRN is minimal, the upper limit values are nearly equal. Therefore, throughout the remainder of this work, we opted to include the CRN in analyses which are too computationally expensive to be completed with both models, such as the sky map analyses described in \autoref{ss:map}. 
\revsjv{We note that we do not find a significantly higher upper limit at any of the frequencies where we found $\mathcal{B}_{10} \sim 10$ for either the CW or CW+CRN models: this indicates that the noise sources are decreasing the posterior PDF at low strain amplitudes but not increasing the posterior PDF at the high strain amplitudes.}
%!H

\newcommand{\htwelve}[1]{1.24}
\newcommand{\heleven}[1]{1.28}

\newcommand{\realtwelve}[1]{1.40}
\newcommand{\realeleven}[1]{1.52}

\newcommand{\rhotwelve}[1]{1.55}
\newcommand{\rhoeleven}[1]{1.22}

\begin{figure}
    \centering
    \includegraphics[width = 1\columnwidth]{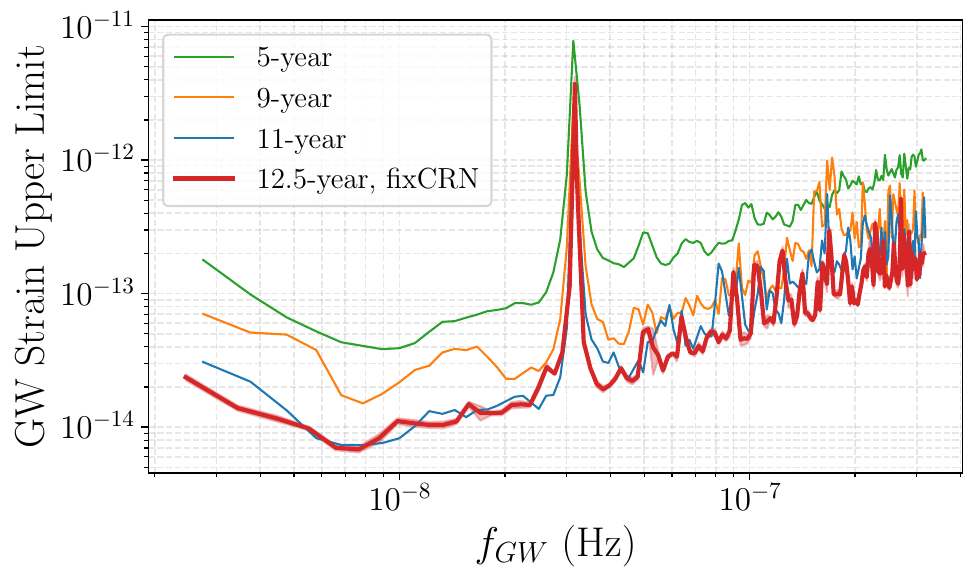}
    \caption{The upper limits on CW strain are continuing to decrease. The 12.5-year data set (red curve and error region) is more sensitive than the 11-year, 9-year, and 5-year (blue, orange, and blue curves, respectively) at high frequencies. At the most sensitive frequency of $f_\mathrm{GW} = $\bestfreq{}, the CRN is impeding further sensitivity improvements, and upper limits are comparable between the 12.5-year and 11-year data sets. At frequencies greater than $f_\mathrm{yr}$, the NANOGrav's sensitivity has improved by a factor of \realtwelve{} since the 11-year data set.}
    \label{fig:ul_compare}
\end{figure}

\revtwo{We are interested in looking at how our sensitivity to CWs changes as we increase the number of pulsars and extend the observing baseline. One approach is to perform ``slice'' analyses, where we truncate the data set to form shorter data sets, and compare the upper limits from the sliced and full data sets \citep{slices}. However, these sliced upper limits are not equivalent to previously published upper limits. Another approach is to directly compare this result to those of previous NANOGrav searches for CWs.}
In \autoref{fig:ul_compare}, we compare this result to those of previous NANOGrav searches for CWs \citep{11yrCW}. \revtwo{Direct comparisons between these data sets are complicated, due to the reprocessing of data resulting in new noise parameter values; nevertheless, such comparisons are useful to examine in order to understand how the sensitivity to CWs improves between data sets.} While analyses have shown a factor of $\sim 2$ improvement between the 
previous three data sets, we see only a modest sensitivity improvement between the 11-year and 12.5-year data, with 
%only a factor of $1.07$ 
\revtwo{a difference of only 7\%}
between the two lowest strain limits \rev{of $h_0 <$ \lowesth{} (12.5-year) and $h_0 <$ \lowesthold{} (11-year)}. In addition to the smaller fractional increase in observing baseline between the 11- and 12.5-year data sets as compared to previous data sets, this is likely due to the presence of the CRN, which, while it is no longer causing a false positive in the CW search if included in the model, does represent a significant noise process that will limit our sensitivity to low-frequency CWs over the years to come \citep{hasasia}.

To confirm this hypothesis, we calculated the sensitivity curves of the 9-, 11-, and 12.5-year data sets using each pulsar's red and white noise contributions and timing model with \texttt{hasasia} \citep{hasasia_code, hasasia} and calculated the relative improvement of in sensitivity between each data set at high frequencies ($>f_\mathrm{yr}$), where red noise has little effect. We observed that on average, the \texttt{hasasia}-calculated sensitivity at these frequencies improved by a factor of \heleven\, between the 9- and 11-year data sets, and \htwelve\, between the 11- and 12.5-year data sets. In our full Bayesian analysis, our upper limits at frequencies above $f_{\mathrm{yr}}$ improved by a factor of \realeleven\, between the 9- and 11-year data sets, and \realtwelve\, between the 11- and 12.5-year data sets. 
%As a second comparison, we used the scaling relation for the signal-to-noise ratio (S/N) for a CW detection in \citet{mingarelli_2017} to compare the theoretical improvement between similar, but lengthened, data sets. There, the S/N $\rho \propto \left<NTc/\sigma^2\right>^{1/2}$, where N is the number of pulsars, T is the timespan of the data set, c is the cadence of observations, and $\sigma$ is the white noise rms. Assuming our pulsars are identical and our observation cadence is unchanged, we can change $N$ and $T$ according to their values in the 9-, 11-, and 12.5-year data sets to compare their relative sensitivity improvement. We calculated that the sensitivity of the data set should increase by a factor of 1.55 between the 9- and 11-year data sets, and a factor of 1.22 between the 11- and 12.5-year data sets. \todo{CONFIRM THIS}
These proportionalities are even greater than our calculated improvements, so we are able to conclude that NANOGrav's sensitivity to CWs is improving as expected at high frequencies where red noise is not dominant.

\vspace{5mm}
\subsection{Sky Map}\label{ss:map}

%The GW strain upper limits, for a model including a CRN, at the most sensitive CW frequency $f_{\rm GW} =~$\bestfreq{} as a function of sky location are shown as a map in \autoref{fig:map}. 
\begin{figure*}
    \centering
    \includegraphics[width = 1.5\columnwidth]{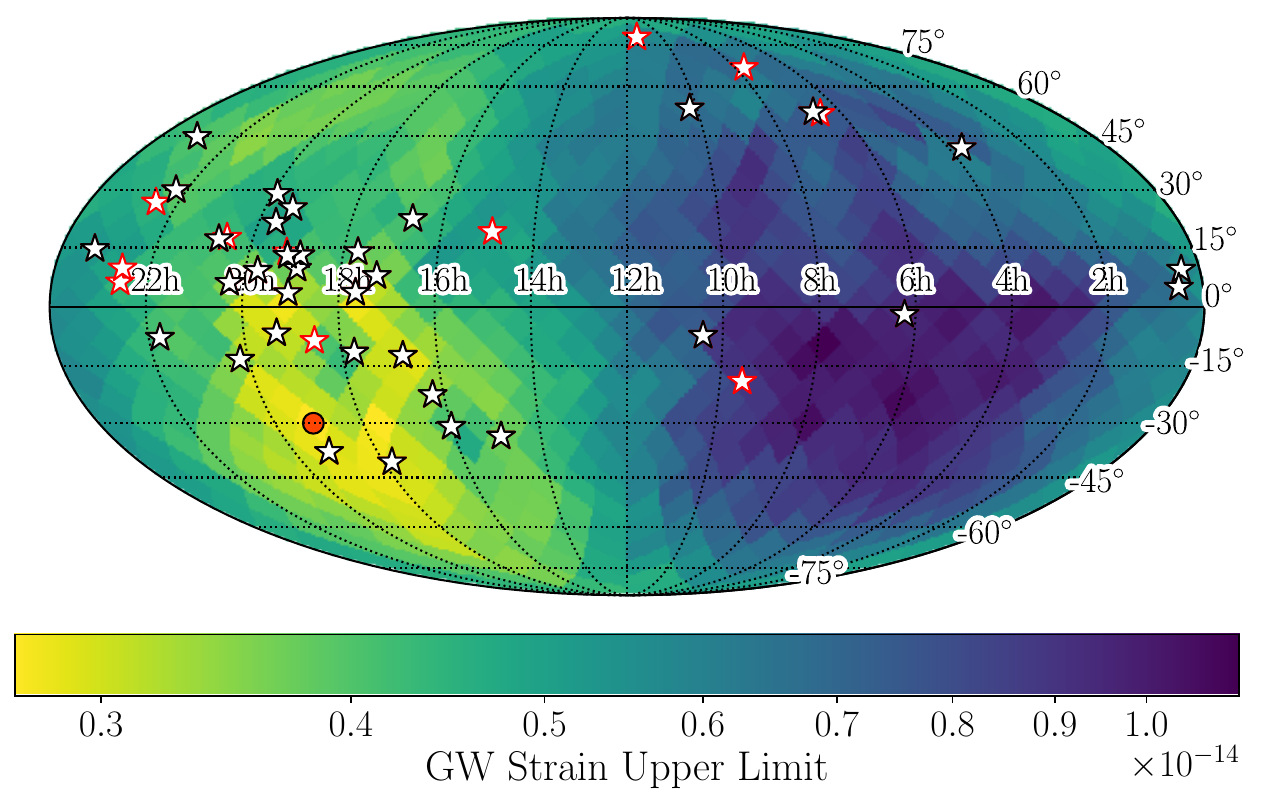}
    \caption{Map of CW strain 95\% upper limits at $f_{\rm GW} = $~\bestfreq{}, the most sensitive frequency searched, for the 12.5-year data set. Pulsar locations are shown as white stars, with new pulsars added from the 12.5-year data set outlined in red. The most sensitive pixel is marked with a red dot, and is located at an RA of $\mathrm{19^h07^m30^s}$ and a Dec of $-30^{\circ}00'00''$. In this region, where the our best-timed pulsars lie, our upper limits are nearly an order of magnitude more sensitive than the least sensitive pixel.}
    \label{fig:map}
\end{figure*}

In \autoref{fig:map}, we show the GW strain upper limits for a model including a CRN at the most sensitive CW frequency $f_{\rm GW} =~$\bestfreq{} as a function of sky location.
As expected, the portion of the sky that is the least sensitive to CWs is that which contains the fewest pulsars. At the most sensitive pixel, the strain upper limit is $h_0 <$ \bestpix{}, while at the least sensitive pixel, $h_0 <$ \worstpix{}, a range of sensitivities that varies by a factor of $\sim 4$.

\begin{figure}
    \centering
    \includegraphics[width = 1\columnwidth]{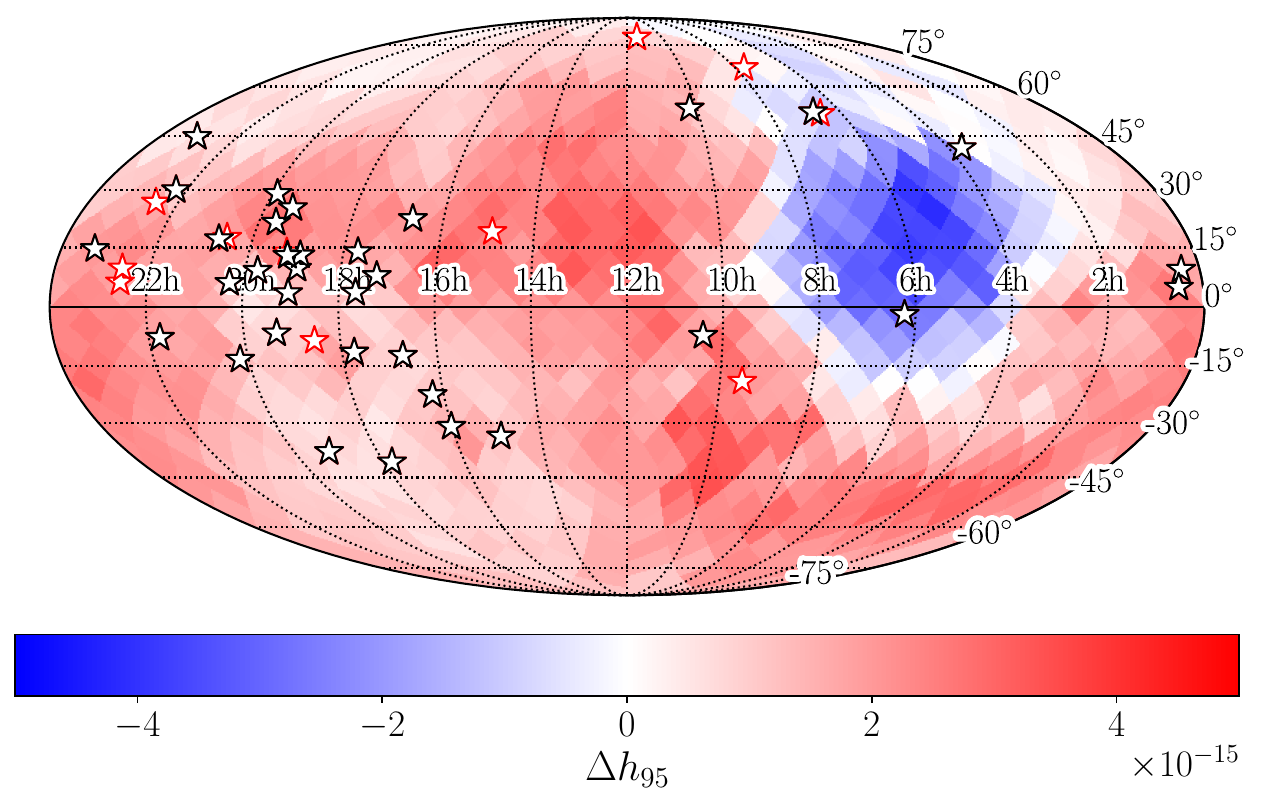}
    \caption{Difference in strain 95\% upper limits for the 12.5-year data set versus the 11-year data set at our most sensitive frequency. Blue pixels indicate a decrease in upper limit, while red pixels indicate an increase. The overall increase in upper limit across much of the sky at the most sensitive frequency was found to be due to the presence of the CRN, and is consistent with the all-sky limit shown in \autoref{fig:ul_compare}.}
    \label{fig:diff_map}
\end{figure}

In \autoref{fig:diff_map}, we compare the 12.5-year CW strain map to that constructed in \citet{11yrCW} for the 11-year data set by plotting $\Delta h_{95} = h_{95, 12.5}- h_{95, 11}$. While a portion \rev{(14\%)} of the sky shows a significant reduction in strain upper limits, \rev{85\% of our pixels} show an increase in strain upper limit, indicating a loss of sensitivity in the newest data set for much of the sky at our most sensitive frequency, including in the most sensitive area of the sky.

\begin{figure}
    \centering
    \includegraphics[width = 1\columnwidth]{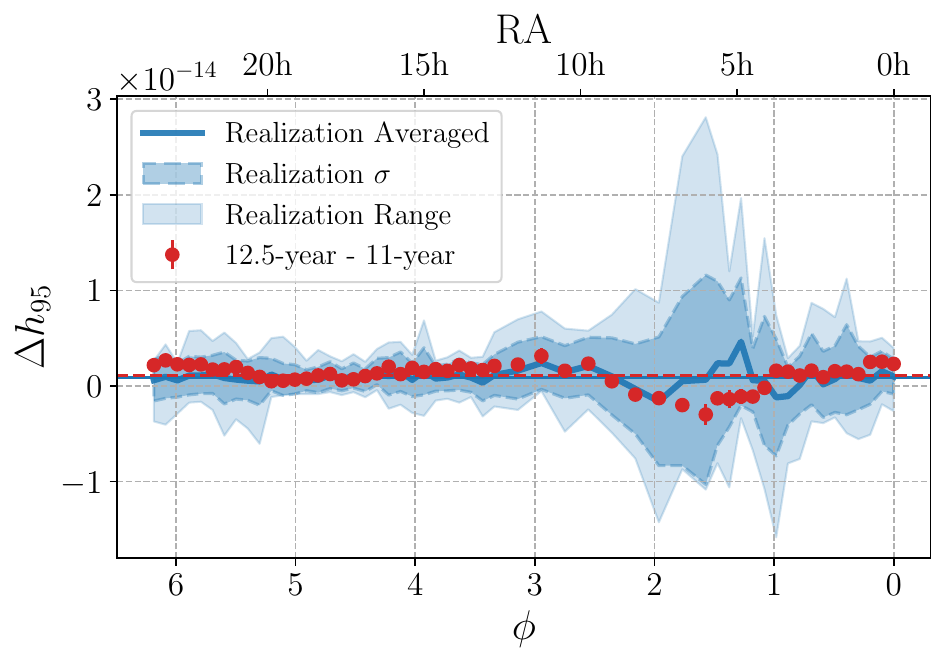}
    \caption{The difference in strain upper limits for an equatorial slice of the sky map shown in \autoref{fig:diff_map} plotted against $\phi$ (or RA). The results for the real data (red points) are well within the range of values encompassed by \rev{the range between and the standard deviation of} ten realizations simulated (blue), with near-identical mean values of $\Delta h_{95}$ (horizontal red and blue lines). Therefore, we conclude that the overall increase in upper limit across much of the sky at our most sensitive frequency is due to the 12.5-year data set's sensitivity to the CRN.}
    \label{fig:sims}
\end{figure}

To investigate the cause of this apparent sensitivity loss, we conducted an analysis of the simulated data utilized in \citet{astro4cast}. We selected portions of the data set with included pulsars and observation baselines corresponding to the 11- and 12.5-year data sets that also included a CRN corresponding to that found in \citet{11yr_gwb}. Then, 
we conducted 
%identical upper limit analyses 
\revtwo{upper limit analyses corresponding to the best-fit model for each data set}
\rev{(i.e. for a CW-only model for the 11-year slice, and a CW+CRN model for the 12.5-year slice)} for an equatorial slice of sky pixels (i.e., for the pixels with $\theta \sim \pi/2$). When plotted against $\phi$ in \autoref{fig:sims}, the patterns in $\Delta h_{95}$ in the real data are well within the range represented by the same analysis in the 10 simulated data sets, each containing a different realization of the CRN. \rev{We observe that between $2h < RA < 8h$, where NANOGrav has the fewest pulsars, the spread of upper limit difference values is by far the largest, which is consistent with our results.} The mean value of $\Delta h_{95}$ across each included pixel is nearly identical for the real data \rev{($\overline{\Delta h_{95}} = 1.12\times 10^{-15}$) and the simulations ($\overline{\Delta h_{95}} = 0.94\times 10^{-15}$), and the real values are well within the intervals spanned by the ten realizations. This indicates that the change in upper limits observed are likely a statistical fluctuation within the range of expected changes shown by our simulations}. Together, this allows us to confidently state that, \revtwo{while exact comparison between data sets is complex,} this apparent pattern in our evolving sensitivity across the sky is due to the emerging CRN. \revtwo{This effect will have significant impacts on future PTA analyses, and will be explored more extensively in future work \citep{simsprep}.}

\section{Astrophysical Limitations of Nearby SMBHBs}\label{sec:astro}

In recent years, numerous studies have modeled the SMBHB population in the nearby universe \citep{simon14, rosado14, schutz, mingarelli_2017, nanograv_nearby} and multiple SMBHB candidates have been discovered with electromagnetic techniques \citep{Sudou2003, Graham2015, spikey, oj287, charisi2016, panstarr}. Even without a CW detection, our limits can add crucial insights into SMBHH populations, including limiting the distance to nearby SMBHBs and placing multi-messenger mass constraints on SMBHB candidates.

\subsection{Distance Limits}
\label{sec:dist}

\begin{figure}
    \centering
    \includegraphics[width = 1\columnwidth]{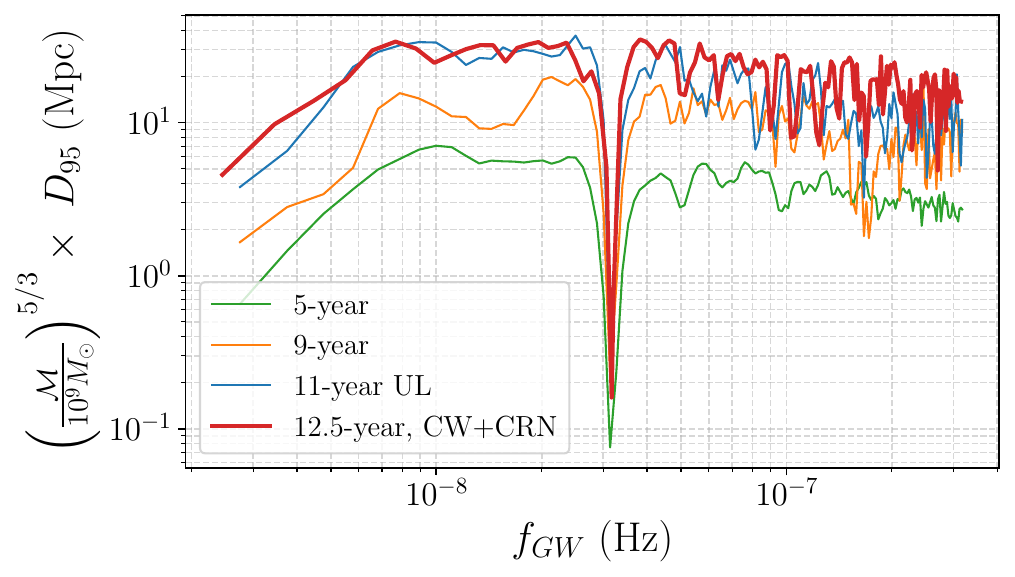}
    \caption{The 95\% lower limits on the luminosity distance to an individual SMBHB. While we can limit SMBHBs emitting GWs at the most sensitive value of $f_{\rm GW} = $ \bestfreq{} to $d_L > $ \senseD{}, at $f_{\rm GW} = $ \farthestF{}, they can be limited to farther away at $d_L > $ \farthestD{}.}
    \label{fig:dist}
\end{figure}

Our limits on CW strain can be transformed using \autoref{eq:strain} to calculate the 95\% lower limit on the luminosity distance to a source of a given chirp mass. The distance limits for an SMBHB with $\mathcal{M} = 10^9 M_\odot$ are shown in \autoref{fig:dist}. For the most sensitive frequency of $f_{\rm GW} = $ \bestfreq{}, we can limit the distance to an SMBHB with $\mathcal{M} = 10^9 M_\odot$ to $d_L > $ \senseD{}. These limits may be scaled to larger or smaller SMBHBs directly using \autoref{eq:strain} as
\begin{equation}
    D_{95, \mathcal{M}} = D_{95, 10^9 M_\odot} \times \left( \frac{\mathcal{M}}{10^9 M_\odot}\right)^{5/3}.
\end{equation}
However, it is important to note that while this frequency produces the lowest strain upper limit, it does not produce the farthest luminosity distance lower limit. This value is $d_L > $ \farthestD{} at $f_{\rm GW} = $ \farthestF{}.

\begin{figure}
    \centering
    \includegraphics[width = 1\columnwidth]{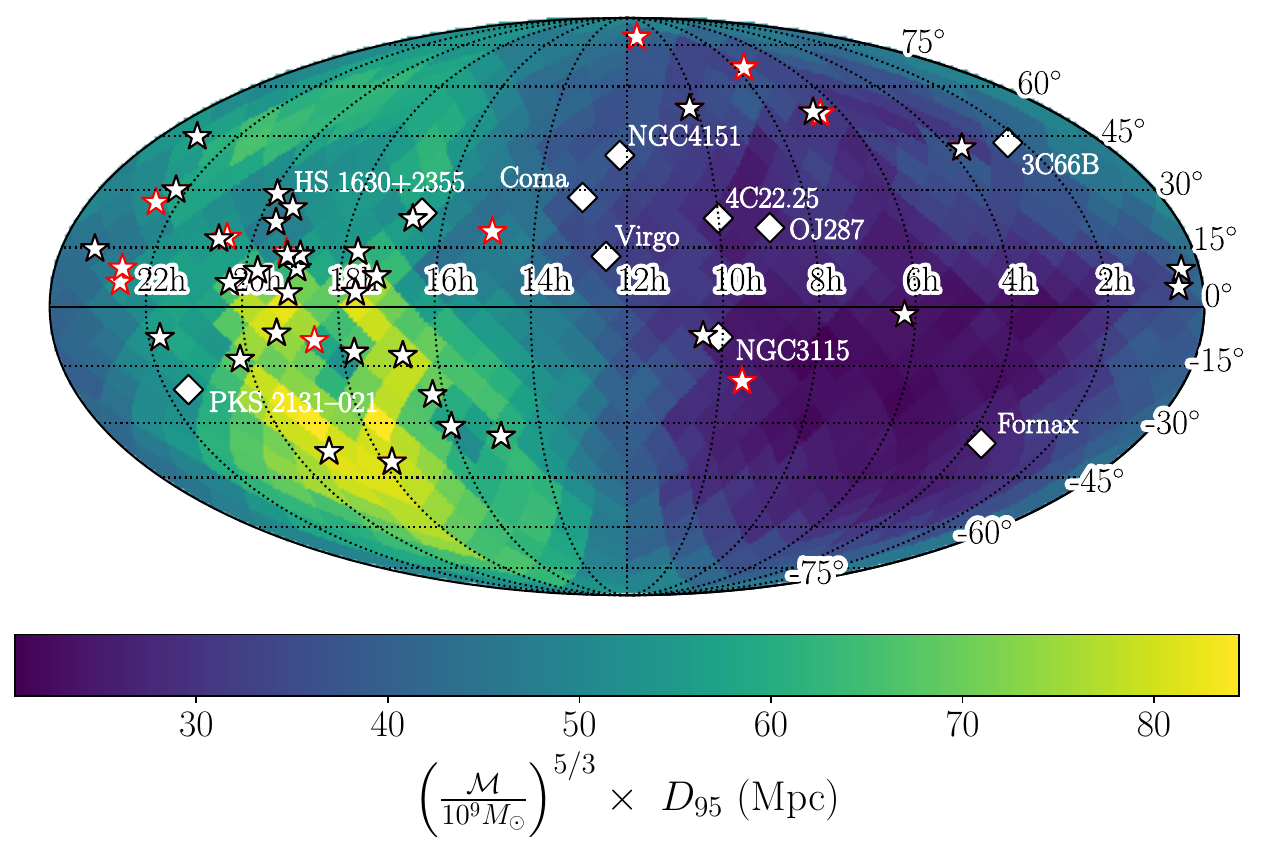}
    \caption{Map of the 95\% lower limit on the distance to individual SMBHBs with $\mathcal{M} = 10^9 M_\odot$ and $7.65\times 10^{-9}$ Hz. White diamonds indicate the positions of known SMBHB candidates and large galaxy clusters that could contain an SMBHB. As PTA sensitivities improve, these candidates may come into reach.}
    \label{fig:dist_map}
\end{figure}

This technique can be applied to the strain upper limit sky map as well, to calculate the 95\% luminosity distance lower limit for an SMBHB emitting CWs at $f_{\rm GW} =~$\bestfreq{} as a function of sky location. The results of this transformation are shown in \autoref{fig:dist_map}. At the most sensitive sky location, we can limit the minimum distance to an $\mathcal{M} = 10^9 M_\odot$ SMBHB to be $d_L >$~\farthestDmap{}, and that to an $\mathcal{M} = 10^{10} M_\odot$ SMBHB to $d_L >$~\farthestDmapTen{}. In the least sensitive sky location, we can limit the minimum distance to an $\mathcal{M} = 10^9 M_\odot$ SMBHB to be $d_L >$~\shortestDmap{}, and that to an $\mathcal{M}=10^{10} M_\odot$ SMBHB to $d_L >$~\shortestDmapTen{}. These values vary by over a factor of 4 between the most and least sensitive parts of the sky.

% At the most sensitive sky pixel, we conducted a final upper limit analysis across the entire frequency band, with results plotted in \autoref{fig:best_loc}. Across the entire nanohertz frequency band, the PTA is dramatically more sensitive to CWs from sources at this sky location than across the entire sky on average. %\todo{say something about the shape once the run is done.}

% \todo{How many realizations have an SMBHB above the curve?}

% \begin{figure}
%     \centering
%     \includegraphics[width = 1\columnwidth]{figures/best_loc_UL.pdf}
%     \caption{
%     %GW frequency and strain for the loudest GW sources for a sample realization of the local Universe, plotted alongside our 95 \% strain upper limit curve. This simulation used simulated populations of nearby SMBHBs from \citep{mingarelli_2017} to determine the number of sources emitting GWs in the PTA band.
%     The 95\% strain upper limit curve for the all-sky (solid red) CW search compared with the 95\% strain upper limit curve in the most sensitive sky location (red dashed).}
%     \label{fig:best_loc}
% \end{figure}
\subsection{SMBHB Number Density Limits}

Using our limits on the luminosity distance to an SMBHB, we can also place limits on the local number density of SMBHBs of a given binary configuration.
After placing a lower limit on the effective comoving distance $d_c$ to sources of given binary parameters, we can say the local density is less than $n_c = 1/V_c = [(4/3) \pi d_c^3]^{-1}$. However, to consider this as a limit on the average density in some volume, that is relatively-local but larger than the explicitly measured volume, there should be some additional pre-factor to account for the confidence of having a source within this volume based on Poisson distributions of sources.
For a number of events 
$\Lambda = n_c V_c$
the likelihood of no detections is $P_0(\Lambda) = e^{- \Lambda}$. 
To find an \rev{upper limit} on the occurrence rate, $\Lambda_\mathrm{UL}$, we must integrate from that limit to infinity, such that the result matches our desired confidence level $p_0$. Therefore,
$F_\mathrm{UL}(\Lambda_\mathrm{UL}) = \int_{\Lambda_\mathrm{UL}}^\infty e^{-\Lambda} d\Lambda = 1 - p_0$ is solved
as 
\begin{equation}
    n_{ul} = \frac{-\ln (1-p_0)}{V_c}.
\end{equation}
Here, our desired confidence level is $p_0 = 0.95$. To calculate the co-moving distance $d_c$, we transform our luminosity distance limits (shown in \autoref{fig:dist}) as $d_c = d_L / (1+z)$, and $z$ is calculated for the relevant luminosity distance values using the \texttt{astropy}.

\begin{figure}
    \centering
    \includegraphics[width = 1\columnwidth]{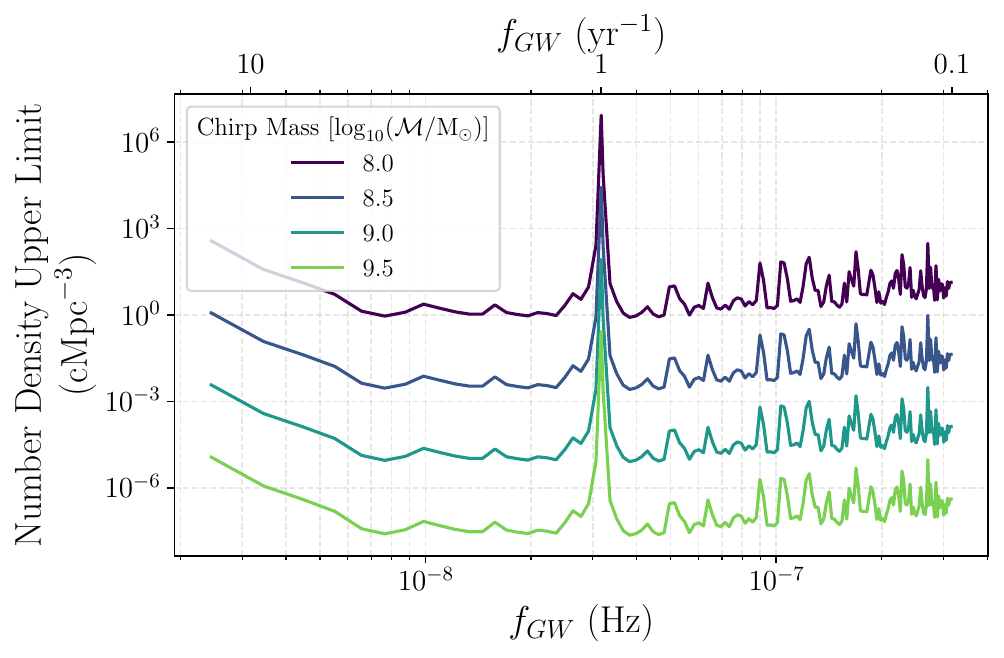}
    \caption{\rev{Number density limits of SMBHBs per comoving Mpc$^{-3}$ in the local universe, where, as expected, we placed significantly more stringent upper limits on the largest SMBHBs than the smallest ones, with limits decreasing from those on chirp masses of $10^8~\mathrm{M_\odot}$ (purple) to $10^{8.5}~\mathrm{M_\odot}$ (blue) to $10^9~\mathrm{M_\odot}$ (teal) to $10^{9.5}~\mathrm{M_\odot}$ (green).}}
    \label{fig:density}
\end{figure}

The results of this calculation are shown for various SMBHB chirp masses in \autoref{fig:density}. As can be expected, we find that we can place more constraining upper limits on large SMBHBs ($\mathcal{M} = 10^{9.5} M_\odot$) than smaller ones ($\mathcal{M} = 10^8 M_\odot$) in the local universe.

\subsection{Multi-Messenger Analyses}

Using the methodology described in \autoref{sec:target_methods}, we conducted a multi-messenger search for GWs from the SMBHB candidate 3C~66B to provide an update to the results of \citep{3c66b}. The detection analyses result in nearly identical Savage-Dickey Bayes factors, whether the CRN was included or not. This is to be expected, as the CRN is very weak at frequencies as high as that of 3C 66B ($f_\mathrm{GW} = 6.04
\times 10^{-8}$ Hz). The Bayes factors for the CW-only analysis and the CW+CRN analysis are \bftargetcw{} and \bftargetcrn{}, respectively. Both of these values are very near 1, meaning that the data do not indicate the presence of a CW corresponding to a binary within 3C~66B. 

\begin{figure}[t]
    \centering
    \includegraphics[width = 1\columnwidth]{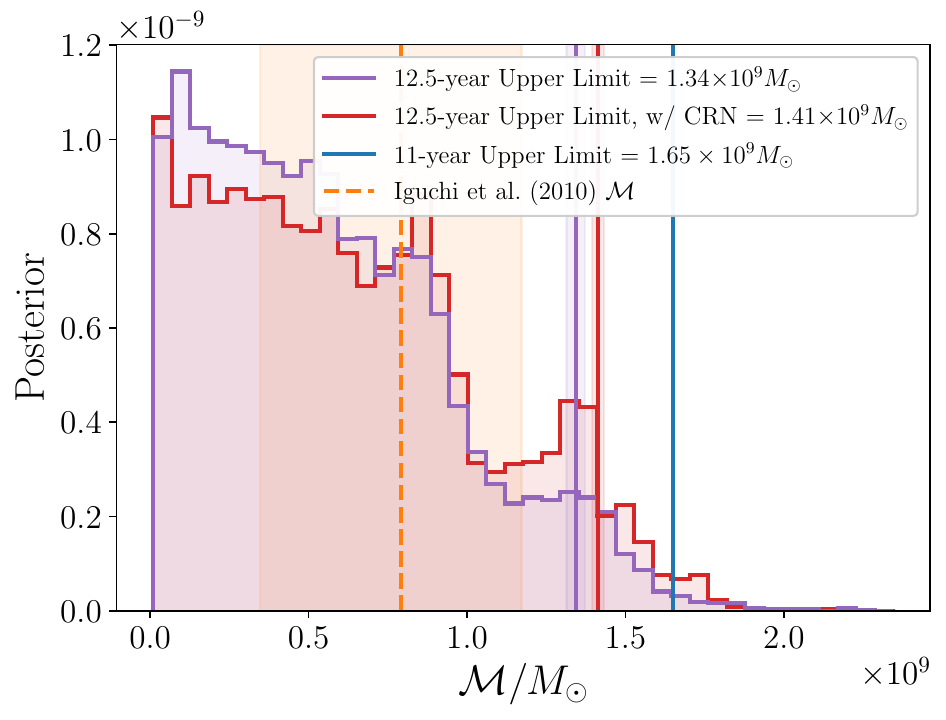}
    \caption{Posterior distributions for a targeted upper limit analysis of the SMBHB candidate 3C 66B. While 95\% upper limits (red and purple lines) are lower than in the 11-year data set (blue line), they cannot rule out the model from \citet{Iguchi2010} (orange region).}
    \label{fig:target}
\end{figure}

Because no GW was detected, we constrain the chirp mass of a potential binary with an upper limit analysis, again performed with and without a CRN to confirm consistency. The posteriors from these two searches are plotted in \autoref{fig:target}, with resulting 95\% upper limits of \ultargetcrn\ when a CRN is included, and \ultargetcw\ when only CWs are included in the signal. For comparison, the 95\% chirp mass upper limit for 3C~66B from the 11-year data set was 
$\mathcal{M} <$ \ultargetold{}. This represents an improvement of \improvement{}, or a factor of \improvementFactor{} smaller; by adding pulsars, extending timing baselines, and improving timing and searching methods, the PTA's sensitivity has clearly improved. These upper limits are nearer to the value of the upper bound of the \citet{Iguchi2010} chirp mass estimate. In subsequent data sets, or by using more sophisticated analyses such as advanced noise modeling \citep{adv_noise_mod}, this error region may soon be within reach.

In \citet{3c66b}, it was shown that a targeted search, like this analysis, results in a factor of $\sim 2$ reduction in upper limits compared to those of an all-sky search at a corresponding GW frequency. When converted to strain amplitudes rather than chirp masses, the 95\% upper limits are $h_0 <$ \ultargetcrnH{} and $h_0 <$ \ultargetcwH{} for the searches with and without a CRN, respectively. In comparison, the all-sky analysis in \autoref{ss:all_sky} returned strain upper limits of $h_0 <$ \ulAllSkyFreqCRN{} and $h_0 <$ \ulAllSkyFreqOnly{} at $\fgw = 6.01
\times 10^{-8}$ Hz, the nearest frequency to that of 3C~66B at \rev{$\fgw = 6.04
\times 10^{-8}$} Hz. These all-sky strain upper limits are a factor of \facCRN{} and \facOnly{} larger, very similar to the value for the 11-year data set. Therefore, the improvement in upper limits gained by using this multi-messenger technique has stayed stable across the addition of new pulsars, more data, and the emergence of the CRN.
% compare to Npsrs and Tobs?
% compare to hasasia SNR?

% \begin{figure}
%     \centering
%     \includegraphics[width = 1\columnwidth]{figures/hs1630_12p5.pdf}
%     \caption{Posterior distributions for a targeted upper limit analysis of the SMBHB candidate HS 1630+2355. The estimated chirp mass for this source from \citet{xin_CRTS} is shown with the blue vertical line.}
%     \label{fig:hs1630}
% \end{figure}

Additionally, we performed a new search for the electromagnetic SMBHB candidate HS 1630+2355. First identified as a periodic quasar in \citet{Graham2015}, this candidate is identified as a top PTA CW candidate in \citet{xin_CRTS} with a gravitational wave frequency of $\fgw = 1.13\times10^{-8}$ Hz and a luminosity distance of 5.26 Gpc. In the 12.5-year data set, we do not detect any CWs from HS 1630+2355; in a CW+CRN analysis (necessary due to the low GW frequency), we calculate a Bayes factor of $0.74 \pm 0.02$. Then, we are able to set an upper limit of \rev{$\mathcal{M}<(1.28\pm0.03)\times 10^{10}M_\odot$} on the chirp mass of an SMBHB within HS 1630+2355, which corresponds to a strain of $h_0 < 4.03 \times 10^{-15}$. For comparison, the all-sky upper limit at the nearest frequency of $\fgw = 1.10\times10^{-8}$ Hz is $h_0 < 1.07\times10^{-14}$, a factor of 2.66 larger than the targeted upper limit. Due to this candidate's favorable position near the PTA's most sensitive sky location, we are able to overcome the much larger source distance to set a constraining upper limit. However, this limit is still approximately 4 times larger than the estimated chirp mass of $3.15\times 10^9 M_\odot$ \citep{xin_CRTS}, meaning that more data are needed to rule out or detect an SMBHB within HS 1630+2355. \rev{The simulations in \citet{xin_CRTS} indicate that HS 1630+2355 will not detectable even by IPTA data sets by the late 2020s, so this result is unsurprising; therefore, HS 1630+2355 will require continued monitoring until PTA sensitivity brings it into reach. }

\subsection{Local Detection Prospects}
\begin{figure}
    \centering
    \includegraphics[width=1\columnwidth]{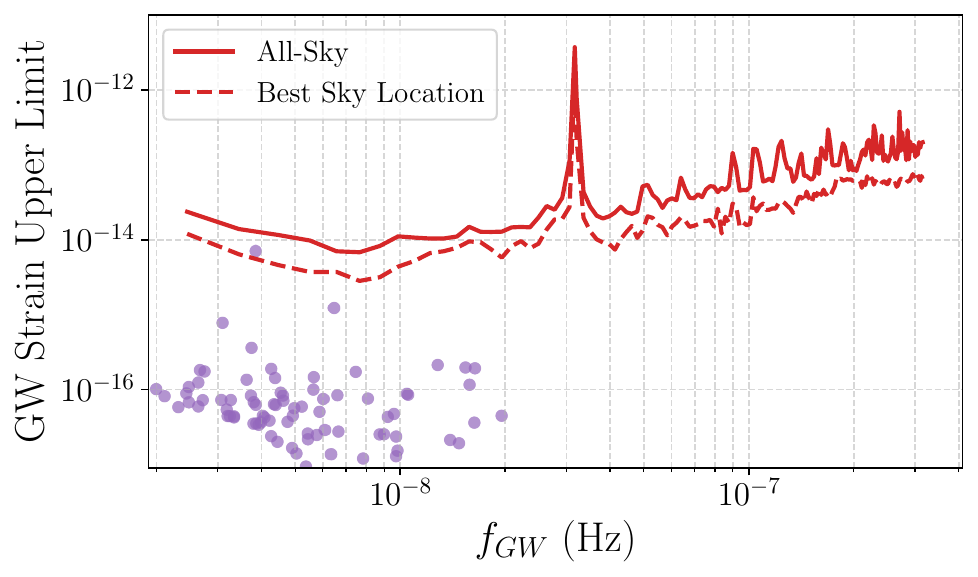}
    \caption{The 95\% strain upper limit curve for the all-sky (solid red) CW search compared with the 95\% strain upper limit curve in the most sensitive sky location (red dashed). The non-detection of a nearby SMBHB is unsurprising -- there was at best a 0.5\% chance of making such a detection. \rev{Here we show one of the 75,000 realizations of the local Universe from \cite{mingarelli_2017}. This realization shows a detectable SMBHB together with our 95\% upper limit curves for both sky-averaged and best sky locations.} In this realization there are 87 local SMBHBs (all within 225 Mpc); none of them lie above the sky-averaged upper limit curve, but one could be detected if it were at the most-sensitive sky location.}
    \label{fig:local_binaries}
\end{figure}

At the most sensitive sky pixel, we conducted a final upper limit analysis across the entire frequency band, with results plotted in \autoref{fig:local_binaries}. Here we observed that for all frequencies, the PTA is dramatically more sensitive to CWs from sources at this sky location than across the entire sky on average. 
\citet{mingarelli_2017} carried out a comprehensive study of the detection prospects of SMBHBs within a 225 Mpc volume, the completeness limit for their chosen K-band luminosity in 2MASS. Using these new 12.5-year upper limit curves, we assess our level of surprise at our current non-detection of CWs. 

\autoref{fig:local_binaries} shows an example realization of the local SMBHB population created with \texttt{nanohertz\_gws} \citep{mingarelli_code}. It is one out of 75,000 Monte Carlo realizations \citet{mingarelli_2017} carried out, where they varied black hole masses via the scatter in various $M-M_\mathrm{bulge}$ relations, mass ratios, and more. While the chosen realization shows what a detectable SMBHB would look like, on average we found only 398 realizations out of the 75,000 contained detectable SMBHB systems at the best sky location. We therefore only had a 0.5\% chance of making a detection of such a local source with the 12.5-year data set. Furthermore, when we consider the entire sky, we found an order of magnitude fewer SMBHBs were detectable -- only 43 realizations contained detectable binaries.

It is interesting to compare this result to that of our previous upper limit \citep{11yrCW}. With the NANOGrav 11-year all-sky upper limits, we found 34 detectable SMBHBs and here we find 43 --- an overall improvement. However, the upper limit at our best sky location has deteriorated due to the CRN, which has in turn decreased the number of detectable binaries by a factor of $\sim 2$, from a $1.2\%$ chance of detection to $0.5\%$.

As was the case in previous sections, we note that this \revtwo{comparison is non-trivial due to the complex changes in sensitivity between data sets}, and that this deterioration is happening primarily at low frequencies where the CRN is manifesting in the data, and the most sensitive sky location is heavily affected (\autoref{fig:diff_map} and \autoref{fig:sims}). \cite{xin_CRTS} show that at higher GW frequencies the effect of the GWB, or any equivalent CRN, is very small, so the detection prospects for local SMBHBs are unaffected.

% \subsection{Frequentist Statistic}
% \todo{fp results}

% \subsection{Massive Binary Detection Prospects}  % alternate section title
\subsection{Binary Population Model Consistency}
\label{sec:pop_model}

\begin{figure}
    \centering
    \includegraphics[width = 1\columnwidth]{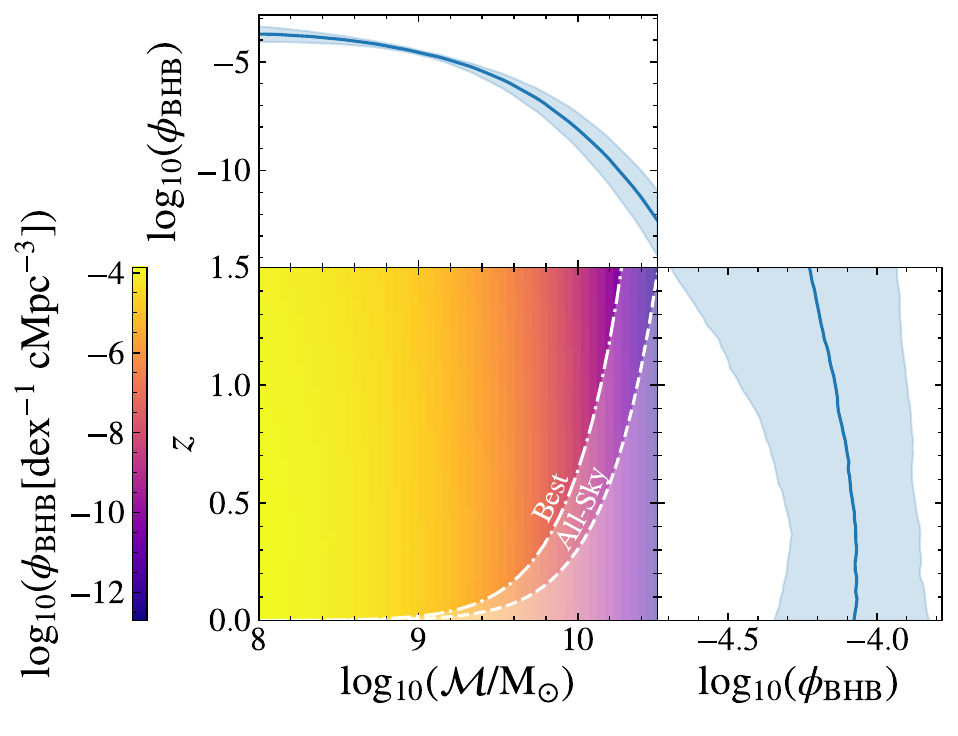}
    \caption{The SMBHB mass function ($\phi_{BHB}$) derived from astrophysical models shows the modeled number density of SMBHBs (color-bar) across log chirp mass ($\mathrm{log_{10}}\mathcal{M}/M_\odot$) and redshift ($z$). Side panels show $\phi_{BHB}$ in one dimension integrated across each respective variable. Regions that are inconsistent with our 12.5-year CW search are shown in white, with the all-sky (average) and most-sensitive (best) sky location upper limits  shown under the solid and dash-dotted white curves, respectively. Created using methods from from \citet{caseyclydefrac}.
    %\todo{update axis formatting}
    }
    \label{fig:population_model}
\end{figure}

Finally, it was also useful to assess whether our current non-detection of CWs is consistent with expectations from SMBHB population models.
In \autoref{fig:population_model} we compared an astrophysically-motivated SMBHB model to GW upper limits set with the 12.5-year CW search.
The SMBHB model was derived from theoretical galaxy major merger rates \citep{chen_constraining_2019}, which are themselves based on observed galaxy pair fractions \citep{mundy} and theoretical galaxy merger timescales.
It is 
%\sout{modeled as in \citet{phinney2001} and \citet{sesana2013} as} 
related to the GWB via \citep{phinney2001,sesana2013}
\begin{equation}
    h_{c}^{2}(f) = \frac{4}{3 \pi}\frac{1}{f^{4 / 3}} \iint \phi_{\rm BHB}(\mathcal{M}, z) \frac{\mathcal{M}^{5 / 3}}{(1 + z)^{1 /3}} d\mathcal{M} dz,
\end{equation}
where $h_c$ is the characteristic strain of the GWB and $\mathcal{M}$ is the chirp mass in the observer frame. This was fit to the results of the NANOGrav search for the GWB in the 12.5-year data set \citep{12p5_gwb}, and assumes the CRN is due to a GWB, comparable to the fit in \citet{middleton_12p5}.

The GW limits in \autoref{fig:population_model} were calculated using the most sensitive frequency of both the all-sky and most-sensitive sky location analyses.
\autoref{fig:population_model} thus shows what regions of $z$--$\mathcal{M}$ parameter space were accessible to the 12.5-year CW search. Since no CWs were detected, we are able to rule out the high-mass and low--$z$ region across the entire sky and at the most sensitive sky location for the PTA's most sensitive frequency.

We calculate the expected number of detectable SMBHBs by relating the differential SMBHB mass function $\phi_\mathrm{BHB}$ to the differential number of binaries per chirp mass, frequency, and redshift \citep{sesana08} as
\begin{equation}
    \frac{d^{3} N}{d \log \mathcal{M} dz df} = \frac{d^{2} \phi_\mathrm{BHB}}{d \log \mathcal{M} dz} \frac{dV}{dz} \frac{dz}{dt_{\rm r}} \frac{dt_{\rm r}}{d f_{\rm r}} \frac{df_{\rm r}}{d f},
\end{equation}
and integrating across the relevant region of $z$ -- $\mathcal{M}$ space, while also considering the entire strain sensitivity curve in frequency space. Here, $t_r$ and $f_r$ are the proper time and binary gravitational wave frequency in the SMBHB's rest frame, respectively.
We find in both cases that the expected number of SMBHBs is $\ll 1$. At the all-sky sensitivity, the calculated number is $0.6 ^{+1.1}_{-0.4}\times 10^{-4}$, while at the most sensitive sky location, the calculated number is $8.6 ^{+12.9}_{-5.5}\times 10^{-4}$.
Our non-detection of a CW signal is thus consistent with theoretical models of the SMBHB population, which predict that the most massive, and therefore loudest, SMBHBs are exceedingly rare.

\section{Discussion and Future Prospects}\label{sec:discussion}
% While the NANOGrav PTA is continuing to add data from ongoing observations, discover new pulsars, and therefore increase our sensitivity to GWs, we have entered an interesting era where surprising results will continue to be uncovered. The presence the CRN first detected in the 12.5-year data set in \citet{12p5_gwb} has impacted the PTA's ability to distinguish a CW source, and therefore, our limitations on CW strains across the nanohertz GW frequency band (\autoref{fig:ul_compare}) and the sky (\autoref{fig:diff_map}) have not improved as steadily as in previous data sets. 

While the NANOGrav PTA is continuing to increase our sensitivity to GWs by adding data from ongoing observations and adding new pulsars to the PTA, our limits on CW strains across the nanohertz GW frequency band and the sky have not improved as steadily as in previous data sets. This is due to the CRN first detected in the 12.5-year data set in \citet{12p5_gwb}, which has impacted the PTA's ability to distinguish a CW source.
While adding a CRN to the search model that is fixed to the maximum-likelihood values from a dedicated search avoids confusion in detection analyses, this adds a significant source of noise to the PTA, and therefore limits our sensitivity to CWs at frequencies below 10 nHz. 

We have entered an interesting era where surprising results will continue to be uncovered. In future data sets, the CRN will likely be even more apparent in the data, and may eventually resolve to be due to a stochastic GWB from SMBHBs \citep{astro4cast}. In any case, due to the multi-frequency nature of the GWB, this will continue to impact CW searches, and significant efforts will be needed to continue development on methods that will allow for efficient detection of both types of nanohertz GW signals such as in \citet{bayeshopper}, as well as extensive simulations that evaluate detection possibilities, as in \citet{astro4cast}, that include multiple types of GW signal in the simulated data sets. Additionally, significant effort will be needed to improve sampling methods that can efficiently explore the complex CW parameter space \citep{quickcw}, particularly at high GW frequencies or if full eccentricity modeling is desired \citep{taylor2016}. \rev{These complexities will only be exacerbated as data sets expand, particularly for the complex data sets produced by the IPTA, which, while more sensitive, contain more pulsars and noise parameters over which to sample.} One promising path forward are targeted searches of quasars, which may be 
%up to $\sim 25$ times 
much more likely to host SMBHBs than random galaxies \citep{caseyclydefrac}. Since multi-messenger analyses can improve upper limits by a factor of 2 \citep{3c66b}, improve detection prospects \citep{liu_mma, charisi_prep}, and can be made drastically more efficient than traditional all-sky searches \citep{charisi_prep}, further development of these methods is also crucial, as with more data, electromagnetic SMBHB candidates may soon be detectable \citep{xin_CRTS}, and many more will be identified in upcoming surveys \citep{charisi_prep,agn_variability}. By balancing these efforts, a CW signal may soon come into reach.

\section{Conclusions}\label{sec:conclusion}
With extensive Bayesian 
%and Frequentist
analyses, we have searched the NANOGrav 12.5-year data set for CWs from individual SMBHBs. In our detection analyses, we showed that no CWs were detected to a high degree of confidence. We then placed all-sky upper limits on the strain amplitude for all CWs emitting between $2.45\times 10^{-9}$ Hz and $3.19\times 10^{-7}$ Hz, as well as upper limits as a function of sky location for the 12.5-year data set's most sensitive frequency of \bestfreq{}. 

This analysis also included the development of new methods to accurately reflect the realistic distribution of possible values of pulsar distances from updated measurements. The way we treat these values in search pipelines has  a significant impact on our ability to detect the pulsar term of a CW signal, and these methods will be critical as we proceed towards PTA sensitivities that enable a CW detection.

Unlike previous data sets, the 12.5-year data set contains a significant CRN. Therefore, for the first time, we included the CRN in our Bayesian searches by fixing the model parameters to those recovered in \citet{12p5_gwb}. This had a significant effect on the results of many of our analyses, and proved critical to avoid a false detection of a CW at $2.45\times 10^{-9}$ Hz. This process also significantly impeded the reduction of our upper limits limits between the 11-year and 12.5-year NANOGrav searches at the most sensitive frequency of \bestfreq{} in most areas of the sky. 
\revsjv{The presence of a CRN will also impact searches for other types of signals, such as bursts with memory and fuzzy dark matter, and so searches for those sources will also need to include the CRN.}

Despite these new necessities, we are able to place significant astrophysical constraints on the local SMBHB population. In our most sensitive sky location, we can rule out the existence of any SMBHB with a mass of at least $10^9 M_\odot$ emitting at \bestfreq{} within \farthestDmap{}. Furthermore, we demonstrate significant improvements to chirp mass upper limits of SMBHB candidates can be made through multi-messenger analysis techniques, and limit the chirp mass of 3C~66B to \ultargetcw{}. With the inclusion of more data, we will soon be able to rule out or confirm this source and other binary candidates, as well as those that are yet undiscovered.

\section{Acknowledgements}\label{sec:acks}
\textit{Author contributions:}
An alphabetical-order author
list was used for this paper in recognition of the fact that a large, decade timescale project such as NANOGrav is necessarily the result of the work of many people. All authors contributed to the activities of the NANOGrav collaboration leading to the work presented here, and reviewed the manuscript, text, and figures prior to the
paper’s submission. Additional specific contributions to this paper are as follows.
ZA, HB, PRB, HTC, MED,
PBD, TD, JAE, RDF, ECF, EF, NG-D, PAG, DCG,
MLJ, MTL, DRL, RSL, JL, MAM, CN, DJN, TTP,
NSP, SMR, KS, IHS, RS, JKS, RS and SJV developed
the 12.5-year data set through a combination of observations, arrival time calculations, data checks and refinements, and timing model development and analysis; additional specific contributions to the data set are
summarized in \citet{12p5_narrow}. CAW coordinated the writing of the paper and led the search. BB, ARK, NSP, JSy, GW, and CAW performed analyses for the project, including exploratory runs. 
JS and CAW developed methods to include the CRN in the search model.
%help jessica and joe have the same initials???
AB, NG-D, JG, KG, SRT, SJV, and CAW proposed for the necessary XSEDE resources to complete these analyses.
NSP and CAW performed the sky map simulations.
AC-C, LZK, CMFM, and CAW developed the astrophysical interpretations. 
ADJ provided updates to red noise empirical distributions.
GEF, XS, and SJV explored the frequentist analyses. 
SC, DJN, MAM, and CAW updated the pulsar distance priors. 
SBS, CMFM, and CAW wrote the manuscript and produced the figures. 
We thank BB, SC, JMC, NJC, WF, KG, JSH, DLK, LZK, MTL, TJWL, MAM, DJN, KDO, JDR, SRT, and SJV for their thoughtful comments on the manuscript.
%rutger

\textit{Acknowledgements.}
This work has been carried out by the NANOGrav collaboration, which receives support
from National Science Foundation (NSF) Physics Frontiers Center award numbers 1430284 and 2020265. The Arecibo Observatory is a facility of the NSF operated under cooperative agreement (No. AST-1744119) by the University of Central Florida (UCF) in alliance with Universidad Ana G. M\'endez (UAGM) and Yang Enterprises (YEI), Inc. The Green Bank Observatory is a facility of the NSF operated under cooperative agreement by Associated Universities, Inc. The National Radio Astronomy Observatory is a facility of the NSF operated under cooperative agreement by Associated Universities, Inc. 
SBS and CAW were supported in this work by
NSF award grant Nos. 1458952 and 1815664. CAW
acknowledges support from West Virginia University
through a STEM Completion Grant, and acknowledges support from CIERA, the
Adler Planetarium, and the Brinson Foundation through a CIERA-Adler postdoctoral fellowship. SBS is a CIFAR Azrieli Global Scholar in the Gravity and the Extreme Universe program. MC and SRT acknowledge support from NSF grant No. AST-2007993. SRT also acknowledges support from an NSF CAREER Award PHY-2146016, and a Vanderbilt University College of Arts \& Science Dean’s Faculty Fellowship. CMFM  was supported in part by the National Science Foundation under Grants NSF PHY-2020265, and AST-2106552. The Flatiron Institute is supported by the Simons Foundation.
Part of this research was carried out at the Jet Propulsion Laboratory, California Institute of Technology, under a contract with the National Aeronautics and Space Administration. Portions of this work performed at NRL were supported by Office of Naval Research 6.1 funding. The Flatiron Institute is supported by the Simons Foundation. Pulsar research at UBC is supported by an NSERC Discovery Grant and by the Canadian Institute for Advanced Research. JS and MV acknowledge support from the JPL RTD program. 
KDO was supported in part by the National Science Foundation under grant No. 2207267. 
ECF is supported by NASA under award number 80GSFC21M0002. 
TD and MTL are supported by an NSF Astronomy and Astrophysics Grant (AAG) award number 2009468. 
LZK was supported by a Cottrell Fellowships Award (No. 27985) from the Research Corporation for Science Advancement made possible by the National Science Foundation grant No. CHE2125978.
MED acknowledges support from the Naval Research Laboratory by NASA under contract S-15633Y.
We acknowledge the use of Thorny Flat at WVU, which
is funded in part by the National Science Foundation Major Research Instrumentation Program (MRI) award No. 1726534 and WVU.
This work used the Extreme Science and Engineering Discovery Environment (XSEDE), which is supported by National Science Foundation grant number ACI-1548562. Specifically, it used the Bridges-2 system, which is supported by NSF award number ACI-1928147, at the Pittsburgh Supercomputing Center (PSC) \citep{bridges}.

\facilities{Arecibo, GBT}

\software{\texttt{enterprise} \citep{enterprise},
\texttt{enterprise\textunderscore extensions} \citep{ee},
\texttt{PTMCMCSampler} \citep{ptmcmc},
\texttt{hasasia} \citep{hasasia_code},
\texttt{libstempo} \citep{libstempo},
\texttt{tempo} \citep{tempo},
\texttt{tempo2} \citep{tempo2},
\texttt{PINT} \citep{pint},
\texttt{matplotlib} \citep{matplotlib},
\texttt{astropy} \citep{astropy, astropy:2013},
\texttt{healpy} \citep{healpy},
\texttt{HEALPix} \citep{healpix},
\texttt{nanohertz\_gws} \citep{mingarelli_code}}
\appendix
\section{Pulsar Distance Values}
\begin{table*}[h]
    \centering
    \caption{Compiled pulsar distance values and uncertainties for each pulsar used in the 12.5-year CW analysis, along with the parallax (PX) or DM prior identifier. Values compiled using measurements from \citet{px1, px2, px3, px4, px5, px6, px7, px8, px9, px10, px11, px12, px13, px14, px15, px16, px17} and \citet{12p5_narrow}.}
    \label{tab:dists}
    \begin{tabular}{cccc|cccc}
    \hline
    Pulsar & Prior & Distance (kpc) & Error (kpc) & Pulsar & Prior & Distance (kpc) & Error (kpc) \\
    \hline
    \hline
B1855+09 & PX & 1.4 & 0.24 & B1937+21 & PX & 3.55 & 0.64 \\
B1953+29 & DM & 4.64 & 0.93 & J0023+0923 & PX & 1.82 & 0.41 \\
J0030+0451 & PX & 0.32 & 0.01 & J0340+4130 & DM & 1.71 & 0.34 \\
J0613-0200 & PX & 1.06 & 0.13 & J0636+5128 & PX & 0.73 & 0.12 \\
J0645+5158 & PX & 1.11 & 0.19 & J0740+6620 & DM & 0.68 & 0.14 \\
J0931-1902 & DM & 1.88 & 0.38 & J1012+5307 & PX & 0.83 & 0.05 \\
J1024-0719 & PX & 1.08 & 0.14 & J1125+7819 & DM & 0.65 & 0.13 \\
J1453+1902 & DM & 1.15 & 0.23 & J1455-3330 & PX & 1.01 & 0.22 \\
J1600-3053 & PX & 1.96 & 0.31 & J1614-2230 & PX & 0.69 & 0.03 \\
J1640+2224 & DM & 1.14 & 0.23 & J1643-1224 & PX & 0.45 & 0.08 \\
J1713+0747 & PX & 1.11 & 0.02 & J1738+0333 & PX & 1.47 & 0.11 \\
J1741+1351 & PX & 2.36 & 0.62 & J1744-1134 & PX & 0.42 & 0.01 \\
J1747-4036 & DM & 3.5 & 0.7 & J1832-0836 & PX & 2.1 & 0.57 \\
J1853+1303 & DM & 2.08 & 0.42 & J1903+0327 & DM & 6.49 & 1.3 \\
J1909-3744 & PX & 1.17 & 0.02 & J1910+1256 & DM & 2.35 & 0.47 \\
J1911+1347 & DM & 2.08 & 0.42 & J1918-0642 & PX & 1.17 & 0.15 \\
J1923+2515 & DM & 1.63 & 0.33 & J1944+0907 & DM & 1.8 & 0.36 \\
J2010-1323 & PX & 2.45 & 0.71 & J2017+0603 & DM & 1.57 & 0.31 \\
J2033+1734 & DM & 1.99 & 0.4 & J2043+1711 & PX & 1.39 & 0.12 \\
J2145-0750 & PX & 0.64 & 0.02 & J2214+3000 & DM & 1.54 & 0.31 \\
J2229+2643 & DM & 1.43 & 0.29 & J2234+0611 & PX & 1.19 & 0.15 \\
J2234+0944 & DM & 1.0 & 0.2 & J2302+4442 & DM & 1.18 & 0.24 \\
J2317+1439 & PX & 1.62 & 0.21 & -- & -- & -- & -- \\
\hline
    \end{tabular}
\end{table*}

\bibliographystyle{aasjournal}
\bibliography{cw}
\end{document}